\documentclass[10pt,letterpaper]{article}
\usepackage[top=0.85in,left=2.75in,footskip=0.75in]{geometry}

\usepackage{amsmath,amssymb}

\usepackage{changepage}

\usepackage{textcomp,marvosym}

\usepackage{cite}

\usepackage{nameref,hyperref}

\usepackage[right]{lineno}

\usepackage{microtype}
\DisableLigatures[f]{encoding = *, family = * }

\usepackage[table]{xcolor}

\usepackage{array}

\usepackage{float}
\usepackage{bbm}
\usepackage{booktabs}
\usepackage{tikz}
\usepackage{multirow}
\usepackage[caption=true]{subfig}

\newcolumntype{+}{!{\vrule width 2pt}}

\newlength\savedwidth

\usepackage[aboveskip=1pt,labelfont=bf,labelsep=period,justification=raggedright,singlelinecheck=off]{caption}

\makeatletter
\renewcommand{\@biblabel}[1]{\quad#1.}
\makeatother

\usepackage{lastpage,fancyhdr,graphicx}
\usepackage{epstopdf}
\fancyheadoffset[L]{2.25in}
\fancyfootoffset[L]{2.25in}
\begin{document}
\vspace*{0.2in}

\begin{flushleft}
{\Large
\textbf\newline{Exploring Workplace Behaviors through Speaking Patterns using Large-scale Multimodal Wearable Recordings: A Study of Healthcare Providers} 
}
\newline
\\
Tiantian Feng\textsuperscript{1},
Shrikanth Narayanan\textsuperscript{1}
\\
\bigskip
\textbf{1} University of Southern California, 
\\ Department of Electrical and Computer Engineering, 
\\ Los Angeles, CA, USC
\\
\bigskip

%
%





* tiantiaf@usc.edu

\end{flushleft}

\section*{Abstract}
Interpersonal spoken communication is central to human interaction and the exchange of information. Such interactive processes involve not only speech and spoken language but also non-verbal cues such as hand gestures, facial expressions, and nonverbal vocalization, that are used to express feelings and provide feedback. These multimodal communication signals carry a variety of information about the people: traits like gender and age as well as about physical and psychological states and behavior. This work uses wearable multimodal sensors to investigate interpersonal communication behaviors focusing on speaking patterns among healthcare providers with a focus on nurses. We analyze longitudinal data collected from $99$ nurses in a large hospital setting over ten weeks. The results indicate that speaking pattern differences across shift schedules and working units. Moreover, results show that speaking patterns combined with physiological measures can be used to predict affect measures and life satisfaction scores. The implementation of this work can be accessed at https://github.com/usc-sail/tiles-audio-arousal.



\section*{Introduction}

Workplace wellness is ever more important, especially in navigating the ever-increasing and competing cognitive, physical, and social demands. Emerging crises such as the COVID-19 pandemic have underscored the importance of investigating ways of supporting the well-being of workers, especially in high-stakes domains such as healthcare. 
Notably, nursing, which is at the core of the modern healthcare system, is facing unprecedented challenges and significant increases in work demand. Nursing is known to be stressful: they are required to work long shifts, typically $12$ hours in duration \cite{richardson_12hour, dabner_12hour}. Increasingly nurses have to handle the intense workload while providing in-patient nursing care that requires constant vigilance and meticulousness. In addition to the physical demands, nursing is also emotionally demanding, requiring the delivery of humane, empathetic, and culturally sensitive care over extended periods \cite{cricco2014need}.
Consequently, they are particularly susceptible to stress, anxiety, and other negative psychological effects that can lead to burnout, job dissatisfaction, and depression \cite{iacovides2003relationship, glass1993depression}. In addition, if not properly managed, high stress and anxiety exacerbate the psychological impact producing further emotional exhaustion, decreased productivity, and, consequently, a more significant challenge for delivering high-quality patient care \cite{circadian_rhythms_boivin, performance_rebecca, nurse_health_costa, letvak_productivity}. Thus, there is a need to investigate working behavior to identify the source of negative impact factors associated with nursing professions.  We investigate workplace behaviors using unobtrusively obtained longitudinal multimodal sensor data in a realworld hospital setting.

Notably, interpersonal communication is fundamental to nursing care in all its facets, such as prevention, treatment, therapy, rehabilitation, education, and health promotion \cite{fakhr2011exploring}. Moreover, scientific evidence has shown that effective interpersonal communication can positively impact quality care and preserve social capital. Studies have also reported that effective interpersonal communication can improve treatment and care plans, enhance the satisfaction of the patients and healthcare professionals, and improve nurses' performance and coordination \cite{leonard2004human}. Interpersonal communication typically includes speech, eye contact, gesture, and digital communication, such as email and text messages. Among them, speech is a primary way for nurses to communicate with others and express themselves during a work shift. Speech signal, in particular, contains a set of objectively measurable parameters that are indicative of the linguistic content communicating intent, the speaker's tone, mental state and health status including stress. Thus, understanding the speech activity of nurses can provide meaningful insights into not only nursing professionals' dynamic state and behavior changes, but how they interact with others in their work environment. 

Two fundamental factors associated with speaking patterns are the frequency and duration of conversational interaction. Scientists have found that frequent communication is critical for developing group cohesiveness, avoiding pileup of work activity, and increasing the trust perceptions among the group \cite{walther2005rules}. Increasing face-to-face communication between health care providers has also been reported to improve the accuracy of care performance \cite{gordon2011unit}. Moreover, emotional cues, such as arousal phenomena (also referred to as activation and excitation) \cite{arousal_berlyne}, from speech signals is another valuable variable to assess speaking patterns \cite{bone2014robust}. The arousal phenomenon is supported by the fact that affective arousal changes involve physiological reactions, which in turn modify the process by which speech is produced \cite{scherer1986vocal}. In general, prosodic features of speech, such as pitch and intensity, can reflect the speaker’s arousal state \cite{bone2012robust, scherer1986vocal}. Arousal, like other affective constructs, is critical to motivation, personality, and work productivity \cite{wilson1990personality, arousal_manage_barsade, johnson1979arousal}. For instance, prior works have studied the association between arousal response and task performance and concluded that arousal is an indicator of personal motivation in task performance and competitions \cite{arousal_manage_barsade, arousal_work_graziotin, malhotra2010desire}. Consequently, understanding the speaking patterns requires the knowledge of how frequent the speech activity is, how long the speech activity last, and how the speaker expresses the speech.

Nurses often work within a hospital unit, which is the micro-organization in the hospital health care system. Units of different types, such as the intensive care unit (ICU) and non-ICU, can create different social cultures and norms. The shift pattern is also a strong factor in impacting the nurse work environment and behaviors. These different work schedules can create substantial differences in how a nurse communicates with others, such as how frequently the speech activity occurs and where the speech activity often occurs. Therefore, we propose to investigate the speech activity with respect to different working schedules, such as the shift pattern and the work environment. In order to quantitatively study speech activity among nursing professionals, the present study uses real-life audio collected from a longitudinal and large-scale study undertaken in the hospital setting. The speech activity and participant proximity to different locations are captured from a novel wearable audio sensor \cite{feng2018tiles}, alongside other multimodal measurement of physiology and activity \cite{mundnich2020tiles2018}. In summary, we aim to \textbf{quantitatively} answer the questions below: \begin{enumerate}
    \item What are the differences in speaking patterns between nurses with different work schedules and work environments?
    \item Do the speaking patterns differ at different locations? 
    \item Are the speaking patterns indicative of a nursing professional's affect and life-satisfaction variables?
\end{enumerate}

\section*{Study Design}
\label{sec:study_design}

In early 2018, we designed and conducted comprehensive experiments in a hospital workplace to examine the physiological, environmental, and behavioral variables affecting hospital employee wellness \cite{mundnich2020tiles2018}. Over ten weeks, we collected data through wearable sensors and surveys from hospital employees working at USC's Keck Hospital in Los Angeles, CA, who had volunteered to be part of this study. All study procedures were conducted in accordance with USC's Health Sciences Campus Institutional Review Board (IRB).

\subsection*{Baseline Assessment}
\label{sec:baseline_variables}
At the beginning of the study, participants were asked to complete sets of baseline assessments related to demographic information, affect level, and life satisfaction. 
\noindent \textbf{Positive and Negative Affect Schedule (PANAS)} was used to assess the affect level \cite{panas_watson}. PANAS consists of $10$ positive and $10$ negative affect items. Participants were asked to rate the extent to which they felt a certain way in the past week on a scale ranging from 1 (\textit{very slightly or not at all}) to 5 (\textit{extremely}). Positive and negative affect scores were calculated by summing individual scores from each group (positive and negative), with higher scores representing higher levels of the corresponding affect.

\noindent\textbf{Life Satisfaction (SWLS)} was a 5-item measure that aims to assess participants' general satisfaction with life. Participants rated the degree to which they agreed with each statement on a scale of 1 (strongly disagree) to 7 (strongly agree). A total score was obtained by taking the average of the 5 items.

\begin{figure*}
    \centering
    \includegraphics[width=\linewidth]{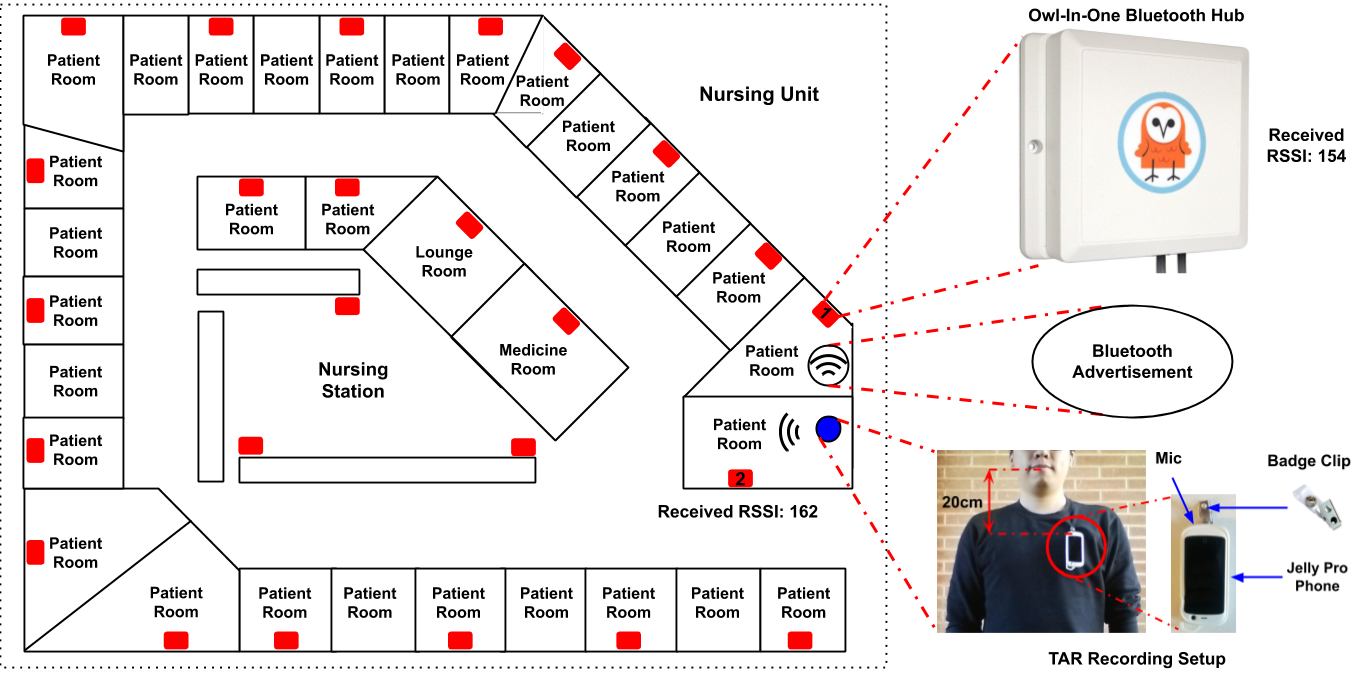}
    \caption{Example of the data recording setup. Red rectangular indicates the location of the Owl-In-One Bluetooth hub in a nursing unit. The Blue circle is a participant in the patient room, and the TAR setup is shown on the right. In this figure, the Own-In-One hub 1 and Own-In-One hub 2 both receive the Bluetooth advertisement from a TAR system where the received RSSI at the Own-In-One hub 1 and the Own-In-One hub 2 are 154 and 162, respectively. }
    \label{fig:layout}
\end{figure*}

\subsection*{Wearable Audio Solution}
Participants were asked to wear a wearable audio sensor called TAR (TILES Audio Recorder) \cite{feng2018tiles} during each work shift throughout the $10$-week period. TAR runs on a small, lightweight, budget-friendly portable Android platform called the Jelly Pro \cite{Jelly}. A complete hardware setup is shown in Fig.~\ref{fig:layout}. In our TAR application, we incorporated the openSMILE feature extractor \cite{openSMILE} to obtain a wide range of features from audio; these audio features have been used to classify emotional states and speaker properties \cite{openSMILE}. TAR uses a lightweight VAD (Voice Activity Detection) module to trigger the feature extraction process only when voice activity is detected to reduce the false negative speech recordings. In our experiment, the VAD runs every 60 seconds and triggers the feature extraction process for 20 seconds when the voice activity is present.

To estimate the location associated with a speech recording, we configured TAR to advertise Bluetooth packets that are picked up by Bluetooth hubs (called owl-in-one) installed throughout the hospital by detecting participant proximity to different locations within each hospital unit. These hubs are mainly located in frequently accessed rooms, including patient rooms, medicine rooms, lounge and break rooms, and nursing stations (computer desks) \cite{booth2019toward}. Figure~\ref{fig:layout} shows an example of the placement of these Bluetooth hubs within a nursing unit. Here, we focus on studying speech activity occurring at the nursing stations and patient rooms as they are the primary working locations for nurses.

\section*{Multi-modal Speaking Pattern Modeling}

\begin{figure*}
    \centering
    \includegraphics[width=\linewidth]{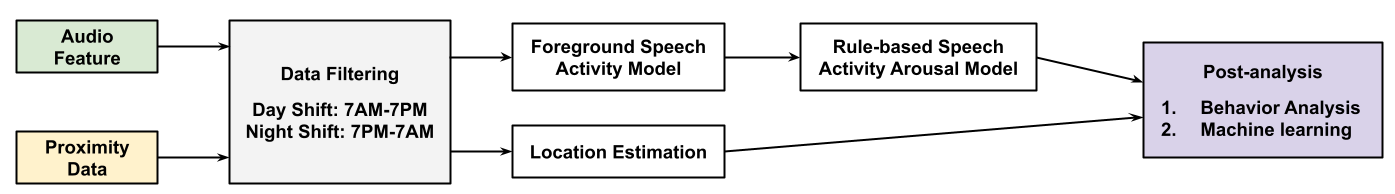}
    \caption{The proposed framework of the in-the-wild Multi-modal Speaking Pattern Modeling.}
    \label{fig:framework}
\end{figure*}

In this section, we describe our framework to extract meaningful speech activity patterns in the multi-modal setting containing indoor proximity information and audio feature samples shown in Figure~\ref{fig:framework}. Our proposed framework starts with filtering the desired data samples based on the nursing work schedules. We then utilize a deep learning model to retain audio feature samples that belongs to the foreground speech activity. Next, we extract speech activity features associated with proximity-based locations to study speaking patterns. A rule-based vocal arousal model is then applied to compute the arousal ratings from the foreground speech features. We also estimate the occurrence of the foreground speech activity from the Bluetooth proximity data. Finally, we conduct a post-analysis combining multi-modal feature streams with learning human behavioral patterns and predicting self-report variables related to affect and well-being.

\label{sec:feat_process}

\subsection*{Data Filtering}
Day and night shift nurses typically work from 7 AM to 7 PM and 7 PM to 7 AM, respectively. Thus, we retain speech activity data and location estimates recorded only in the corresponding shift schedule. We further keep participants with a minimum of 5 days of recordings for the following analysis.

\subsection*{Location Estimation}

Proximity is measured using the Received Signal Strength Indicator (RSSI) each time the hubs receive the Bluetooth packet from TAR. In this study, the RSSI values reported by the hubs range from 136 (i.e., distant) to 193 (i.e., proximal). As RSSI readings can be noisy with distance increasing, we decide to remove RSSI measurements below under 150. We estimate the type of location a participant stayed in at an effective frame rate of one estimate per minute. In particular, we use the highest RSSI from all available observations in a minute to estimate the participant's location. In summary, we investigate the speech activity associated with $5$ location types: \textbf{1. All location}, which includes the patient room, the nursing station, the lounge, the medicine room, and outside the unit; \textbf{2. The nursing station (NS)}; \textbf{3. The patient room (Pat)}; \textbf{4. The lounge and medicine room (Lounge+Med.)}; \textbf{5. Outside the unit}. We chose to combine the speech activity in the medicine room and the lounge due to the limited number of samples recorded at each location.

\subsection*{Foreground Speech Activity}

\begin{figure}
    \centering
    \includegraphics[width=0.8\linewidth]{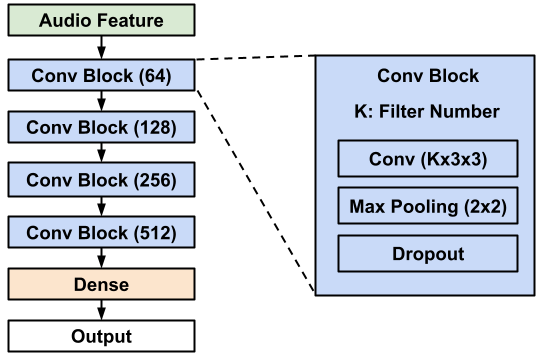}
    \caption{Architecture of the Foreground Speech Activity Detection model.}
    \label{fig:model}
\end{figure}

One primary concern with recording audio in the wild is that, apart from the participant's speech signal, the recordings may also contain background acoustic noise, cross-talk from other people, etc. This can significantly impact the reliability of post-analysis. Therefore, it is required to retain the speech activity data only belonging to the participant of interest. We perform this filtering of the features to obtain the participant's speech feature samples using the machine learning model proposed in \cite{nadarajan2019speaker}. The detailed model architecture is shown in Figure~\ref{fig:model}. The model consists of 4 convolutional blocks following a dense layer to predict whether an audio frame is from the foreground speech activity. This model was trained using a publicly available meeting corpus \cite{janin2003icsi}. Further, model adaptation is performed using a different corpus collected in-house using the TAR system. Finally, we keep the foreground vocal features in the following analysis.

\subsection*{Speaking Pattern Extraction}

This subsection presents the method to calculate features that describe speech activity. First, we aggregate (by averaging) the speech activity features from each participant from all recorded shifts. Then, to study the dynamics of speech activity in a work shift, we divide each work shift into 12 uniform time blocks to investigate how the speech activity varies across a work shift. 

\subsubsection*{Speech Session}

\begin{figure}
    \centering
    \includegraphics[width=0.95\linewidth]{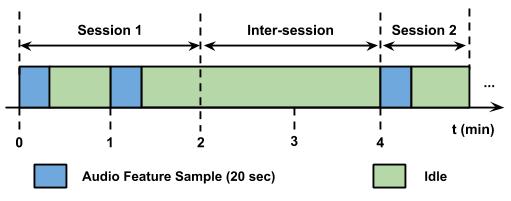}
    \caption{Example of the speech activity sessions. The x-axis represents the time in minutes. The first session and the second session contain the speech activity at $t={0, 1}$ and $t={4}$, respectively. The session $1$ is a $\mathbf{>1}$\textbf{min session} with a session time of $2$ minutes, while the session $2$ is a $1$min session. \textbf{The inter-session time} between the session $1$ and the session $2$ is $2$ minutes.}
    \label{fig:session_def}
\end{figure}

The speech recordings are sampled discretely in time since TAR uses a VAD module to trigger the speech feature recording. Thus, we define a \textit{speech session} as a set of discrete speech recordings with an offset of one minute between each consecutive recording (the recording also must have at least $200$ foreground speech frames). Figure \ref{fig:session_def} shows an example speech activity session, where a set of speech recordings are at time $\mathbf{t}=\{0, 1, 4\}$, and we can group these speech recordings into two sessions at a time $\mathbf{t_{1}}=\{0, 1\}$ and at time $\mathbf{t_{2}}=\{4\}$. 

\vspace{1mm}
\noindent \textbf{Inter-session time} is the time interval between two consecutive speech sessions. This feature describes how frequently a nurse interacts with others. 
\vspace{1mm}

\noindent \textbf{$\mathbf{>}$1min session ratio} is the number of sessions above 1 minute divided by the total number of sessions. This feature describes how likely a nurse is to engage in lengthy continuous speech activity.
\vspace{1mm}

\noindent \textbf{Session occurrence rate (location)} is the ratio between the total session time associated with a location category and the total session time. We use this feature to study relations between speech activity and indoor locations.
\vspace{1mm}

\subsubsection*{Speech Arousal}

In this study, we apply a rule-based framework to estimate the arousal expressed via speech, as proposed in \cite{bone2014robust}. This method has been validated to achieve a strong correlation and classification accuracy between the estimated vocal arousal scores and ground-truth arousal rating in diverse scenarios. The solution is generalizable and favorable for cases where collecting raw training samples is not viable. This method only requires small sets of acoustic features but not labeled emotional training data. The selected feature sets used in this study are log-pitch, intensity, and high-frequency energy to low-frequency energy ratio. First, this framework builds neutral (baseline) models $N_{i}$ for each feature type $i\in\{1, 2, 3\}$. As the authors in \cite{bone2014robust} suggested, we aggregate all the data for a particular speaker to construct the baseline model. Then, the arousal intensity score $p_{i, j}$ is computed as the median value of feature type $i$ in the $j$-th recording with the corresponding neutral model, $N_{i}$, by:

\begin{equation}
    p_{i, j} = 2 \times E[x_{i, j} > N_{i}] - 1
    \label{equ:arousal}
\end{equation}

Then a fusion technique is applied to combine the scores from each feature into a single arousal rating. The weights for fusion are calculated as the Spearman's rank-correlation coefficient $r_{i}$ between each score vector $p_{i}$ and the score mean vector $p_{\mu}$, where the vectors are composed of scores for all of a speaker's utterances. We define the correlation vector as $\mathbf{r} = (r_{1}, r_{2}, r_{3})$, and the weights are then normalized to have a combined magnitude of 1 as shown below:

\begin{equation}
    w_{i} = \frac{r_{i}}{||\mathbf{r}||}
    \label{equ:weight}
\end{equation}

The authors in \cite{bone2014robust} suggest that if annotated neutral affect baselines are established, the obtained arousal scores may be interpreted such that positive (negative) scores indicated positive (negative) arousal, with the magnitude being associated with confidence values. However, since we do not have an annotated neutral baseline, we can instead interpret the rated positive (negative) scores 
as positive (negative) compared to the participant's baseline data. Thus, we define the baseline arousal ratings at 0.25 and -0.25 as the threshold for positive and negative arousal speech for each participant, respectively. Finally, we can define the \textbf{positive (negative) arousal speech ratio} as the number of speech activities above (below) the positive (negative) arousal threshold divided by the total number of speech activities. Here, positive arousal can be related to happiness, nervousness, anger, etc., while negative arousal can be regarded as calm, relaxed, bored, etc.

\section*{Baseline Assessment Results}

\subsection*{Demographic}

The sample included $n = 99$ nurses who are with a minimum of 5 days of recordings, of which $n = 56$ ($56.6\%$) worked the day shift, and $n = 33$ ($33.8\%$) worked the night shift. The majority of participants are female ($n = 70$, $70.7\%$), and all of them have a college degree. The age of participants ranges from $25$ -- $65$, with $13.1\%$ of participants $20$-$29$ years old, $48.5\%$ of participants $30$-$39$ years old, and $38.4\%$ of participants $40$ years old. The average number of years spent working as a nurse was $10.54$ years (SD = $8.66$); of which participants took the management questions, and $n = 37$ ($37.4\%$) of participants managed other nurses.

\begin{table}
    \centering
    \caption{Table showing the $>$1min session ratio (denoted as $>$1min session \%), inter-session time (in minutes), the session occurrence rate (location), and positive/negative arousal speech activity ratio between two work schedules. Statistical significance is performed using the Mann–Whitney U test: $\mathbf{p^{*}<0.05}$. The median and the mean value of each feature are presented with the notation: median (mean).}
    \vspace{1mm}
    \small
    \begin{tabular}{p{2cm}p{1.5cm}p{1.5cm}p{1.5cm}}

        \toprule
        \multicolumn{1}{l}{\textbf{Feature}} &
        \multicolumn{1}{c}{\textbf{Day shift (n=56)}} &
        \multicolumn{1}{c}{\textbf{Night shift (n=43)}} &
        \multicolumn{1}{c}{\textbf{p-value}} \rule{0pt}{2.25ex} \\
        
        \cmidrule(lr){1-1} \cmidrule(lr){2-3} \cmidrule(lr){4-4}
        
        \multicolumn{1}{l}{\textbf{Inter-session Time}} &
        \multicolumn{1}{c}{$5.79$ ($6.71$)} &
        \multicolumn{1}{c}{$8.53$ ($9.86$)} &
        \multicolumn{1}{c}{$\mathbf{<0.01^{*}}$} \rule{0pt}{2.25ex} \\

        \multicolumn{1}{l}{\textbf{$\mathbf{>}$1min Session \%}} & & &  \rule{0pt}{2.25ex} \\

        \multicolumn{1}{l}{\hspace{0.25cm}{All Location}} &
        \multicolumn{1}{c}{$38.04$ ($36.57$)} &
        \multicolumn{1}{c}{$30.57$ ($30.77$)} &
        \multicolumn{1}{c}{$\mathbf{<0.01^{*}}$} \rule{0pt}{2.25ex} \\
        
        \multicolumn{1}{l}{\hspace{0.25cm}{Nursing Station}} &
        \multicolumn{1}{c}{$30.52$ ($29.50$)} &
        \multicolumn{1}{c}{$22.44$ ($23.37$)} &
        \multicolumn{1}{c}{$\mathbf{<0.01^{*}}$} \rule{0pt}{2.25ex} \\
        
        \multicolumn{1}{l}{\hspace{0.25cm}{Patient Room}} &
        \multicolumn{1}{c}{$34.10$ ($32.26$)} &
        \multicolumn{1}{c}{$26.71$ ($27.17$)} &
        \multicolumn{1}{c}{$\mathbf{0.019^*}$} \rule{0pt}{2.25ex} \\
                
        \multicolumn{1}{l}{\textbf{Session Occ. \%}} & & &  \\
        
        \multicolumn{1}{l}{\hspace{0.25cm}{Nursing Station}} &
        \multicolumn{1}{c}{$30.40$ ($31.11$)} &
        \multicolumn{1}{c}{$36.84$ ($36.62$)} &
        \multicolumn{1}{c}{$\mathbf{0.037^*}$} \rule{0pt}{2.25ex} \\
        
        \multicolumn{1}{l}{\hspace{0.25cm}{Patient Room}} &
        \multicolumn{1}{c}{$43.40$ ($43.05$)} &
        \multicolumn{1}{c}{$47.43$ ($46.72$)} &
        \multicolumn{1}{c}{$0.112$} \rule{0pt}{2.25ex} \\
        
        \multicolumn{1}{l}{\hspace{0.25cm}{Lounge+Med.}} &
        \multicolumn{1}{c}{$11.04$ ($12.61$)} &
        \multicolumn{1}{c}{$7.71$ ($8.36$)} &
        \multicolumn{1}{c}{$\mathbf{<0.01^{*}}$} \rule{0pt}{2.25ex} \\
        
        \multicolumn{1}{l}{\hspace{0.25cm}{Outside the Unit}} &
        \multicolumn{1}{c}{$12.21$ ($14.20$)} &
        \multicolumn{1}{c}{$8.47$ ($9.91$)} &
        \multicolumn{1}{c}{$\mathbf{<0.01^{*}}$} \rule{0pt}{2.25ex} \\

        \multicolumn{1}{l}{\textbf{Pos. Arousal \%}} & & &  \\
        
        \multicolumn{1}{l}{\hspace{0.25cm}{All Location}} &
        \multicolumn{1}{c}{$24.48$ ($25.19$)} &
        \multicolumn{1}{c}{$25.11$ ($24.78$)} &
        \multicolumn{1}{c}{$0.199$} \rule{0pt}{2.25ex} \\
        
        \multicolumn{1}{l}{\hspace{0.25cm}{Nursing Station}} &
        \multicolumn{1}{c}{$18.23$ ($20.41$)} &
        \multicolumn{1}{c}{$18.46$ ($19.15$)} &
        \multicolumn{1}{c}{$0.386$} \rule{0pt}{2.25ex} \\
        
        \multicolumn{1}{l}{\hspace{0.25cm}{Patient Room}} &
        \multicolumn{1}{c}{$28.39$ ($28.51$)} &
        \multicolumn{1}{c}{$27.37$ ($29.18$)} &
        \multicolumn{1}{c}{$0.194$} \rule{0pt}{2.25ex} \\

        \multicolumn{1}{l}{\textbf{Neg. Arousal \%}} & & &  \\
        
        \multicolumn{1}{l}{\hspace{0.25cm}{All Location}} &
        \multicolumn{1}{c}{$26.57$ ($26.88$)} &
        \multicolumn{1}{c}{$28.20$ ($31.19$)} &
        \multicolumn{1}{c}{$\mathbf{0.023^*}$} \rule{0pt}{2.25ex} \\
        
        \multicolumn{1}{l}{\hspace{0.25cm}{Nursing Station}} &
        \multicolumn{1}{c}{$30.29$ ($29.42$)} &
        \multicolumn{1}{c}{$33.36$ ($36.08$)} &
        \multicolumn{1}{c}{$\mathbf{<0.01^{*}}$} \rule{0pt}{2.25ex} \\
        
        \multicolumn{1}{l}{\hspace{0.25cm}{Patient Room}} &
        \multicolumn{1}{c}{$20.86$ ($22.39$)} &
        \multicolumn{1}{c}{$23.61$ ($25.62$)} &
        \multicolumn{1}{c}{$0.076$} \rule{0pt}{2.25ex} \\

        \bottomrule

    \end{tabular}
    
    \label{tab:session_shift}
    \vspace{-3.5mm}
\end{table}

\begin{figure}
    \centering
    \begin{tikzpicture}
        
        \node[draw=none,fill=none] at (0, 3.5){\includegraphics[width=0.8\linewidth]{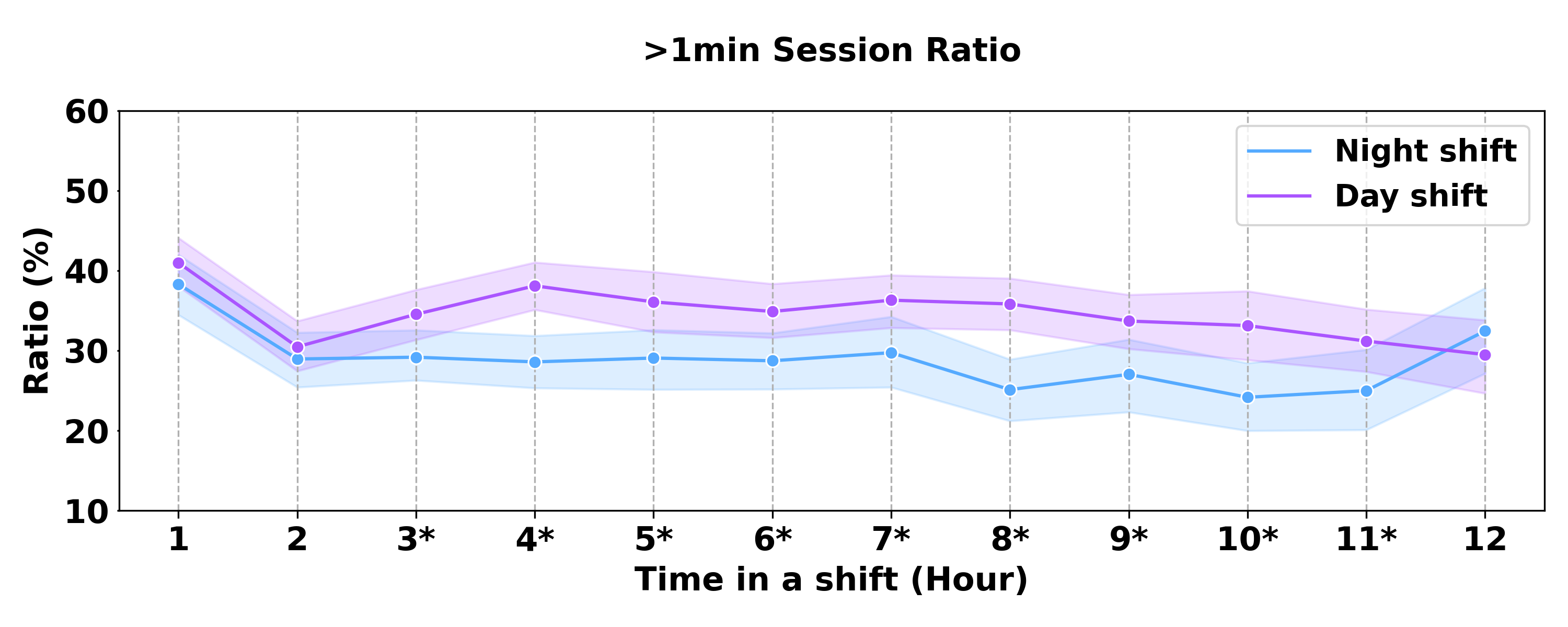}};
        
        \node[draw=none,fill=none] at (0, 7.5){\includegraphics[width=0.8\linewidth]{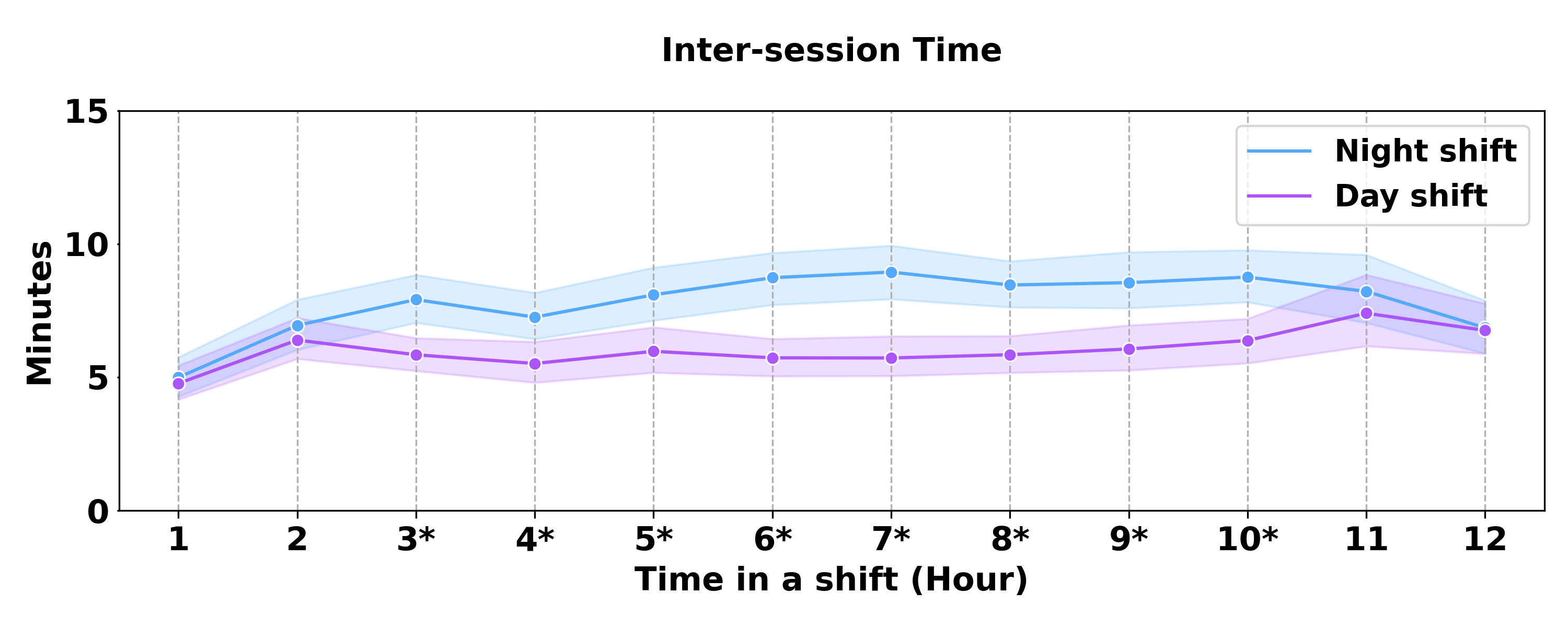}};

    \end{tikzpicture}
    
    \vspace{-3.5mm}
    
    \caption{Comparison of the inter-session time (in minutes) and $>$1min session ratio (\%) between day shift and night shift nurses. Statistical significance is performed using the Mann–Whitney U test: $\mathbf{p^{*}<0.05}$.}
    \label{fig:day_night_session}
    
\end{figure}

\begin{figure}
    \centering
    \begin{tikzpicture}
        
        \node[draw=none,fill=none] at (0, 3.5){\includegraphics[width=0.8\linewidth]{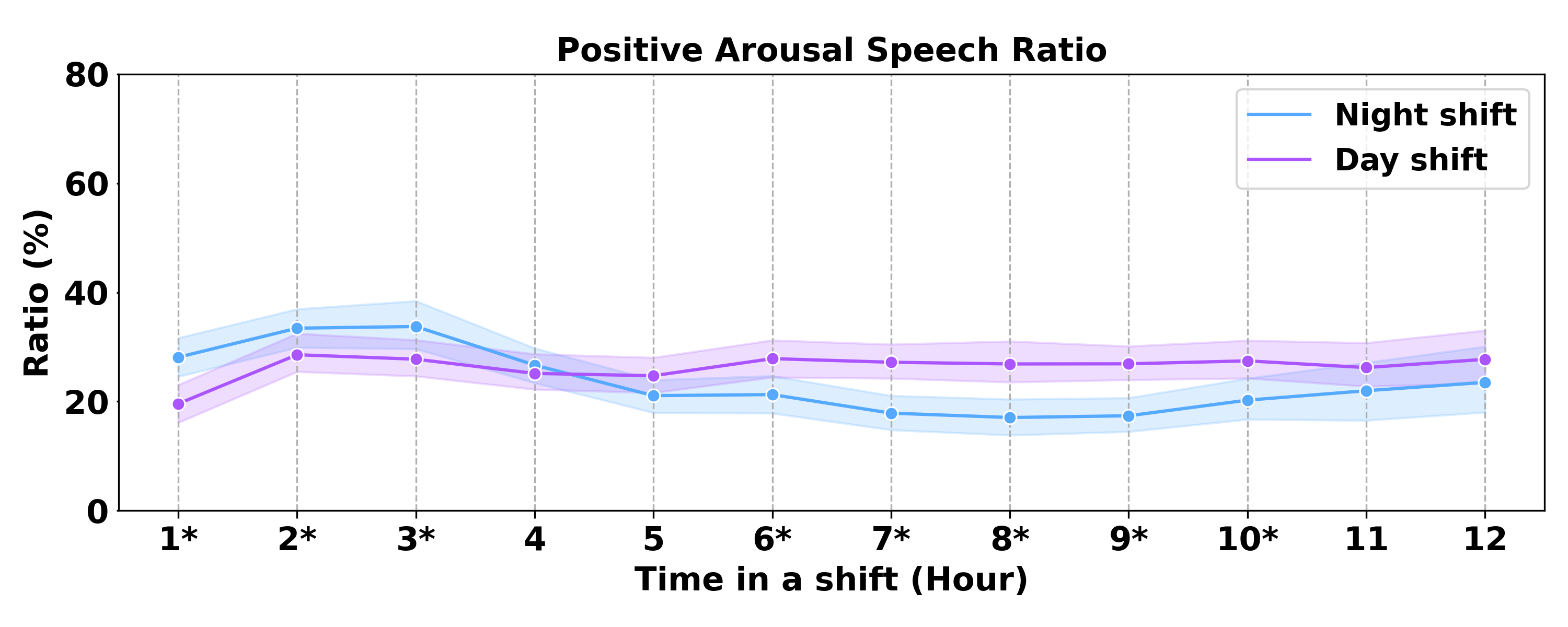}};
        
        \node[draw=none,fill=none] at (0, 7.5){\includegraphics[width=0.8\linewidth]{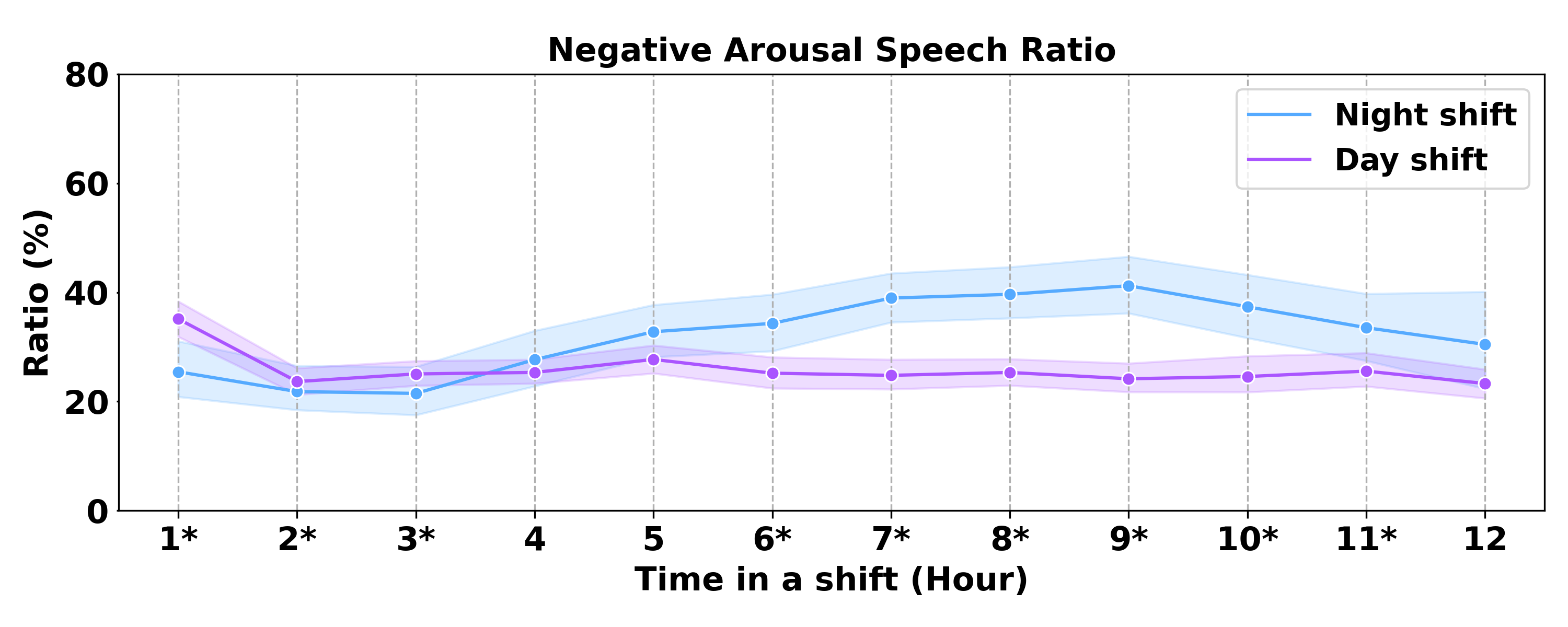}};

    \end{tikzpicture}
    
    \vspace{-2.5mm}
    
    \caption{Comparison of the positive and negative arousal speech activity ratio (\%) between day shift and night shift nurses. Statistical significance is performed using the Mann–Whitney U test: $\mathbf{p^{*}<0.05}$.}
    \label{fig:day_night_arousal}
    
\end{figure}

\section*{Speaking Patterns: Relation to Shift schedule}

\subsection*{Speech Session}

\noindent \textbf{Statistic Analysis} Here, we first present the comparisons of the inter-session time and $>$1min session ratio between day shift nurses and night shift nurses. From the table \ref{tab:session_shift}, day shift nurses have a significantly lower inter-session time than night shift nurses ($p<0.01$). This finding indicates that day shift nurses have a much higher rate of communicating with others than night shift nurses during work. From the comparisons related to the $>$1min session ratio, we can also find that day shift nurses tend to have lengthier interactions at the nursing station (median=30.40\%, mean=31.11\%). Moreover, we observe that nurses, regardless of shift patterns, have more interpersonal communications in the patient rooms than at the nursing station. We also discover that night shift nurses have a higher speech session occurrence ratio than day shift nurses in the nursing station, lounge+medicine room, and outside the working unit.

\vspace{1mm}
\noindent \textbf{Dynamic Patterns} To investigate how the speech activity session features vary at different periods in a shift, we further plot the inter-session time and $>$1min session ratio at different times in a work shift in Figure~\ref{fig:day_night_session}. We select these 2 features as they are indicative of the shift pattern. Overall, we see that nurses who work a day shift interact more frequently with others than nurses who work a night shift between $3^{rd}$ and $11^{th}$ hour in a shift. Nevertheless, day shift nurses consistently have longer speech activity sessions than night shift nurses throughout a work shift. Next, we observe that day shift nurses working in day time have a more diverse inter-session time at the nursing station towards the end of the shift, while the same behavior occurs in ICU nurses who work a night shift.

\subsection*{Speech Arousal Patterns}

\noindent \textbf{Statistical Analysis:} In this subsection, we present the analysis of the arousal rating from the recorded speech data using the rule-based framework described in Section~\ref{sec:feat_process}. Similar to the analysis of the speech activity sessions, we compare the positive and the negative arousal speech ratio between nurses who work a day shift and those who work a night shift. In addition, we include the comparisons in the nursing station and patient rooms as speech activity occurs most in these two locations. From the comparisons shown in Table~\ref{tab:session_shift}, we identify that night shift nurses express more negative arousal in the speech activity, particularly in the nursing station and outside the working unit. However, there are no differences in the arousal expressed from speech activity in the patient room area.

\noindent \textbf{Dynamic Patterns:} The dynamic changes of the arousal expressed from the speech activity between two shift patterns are presented in Figure~\ref{fig:day_night_arousal}. Despite there being no differences in the negative arousal speech ratio between the two shift schedules, we can recognize that night shift nurse engages in more speech activity with positive arousal at the start of a work shift. However, this positive ratio score decreases significantly in the middle of a work shift for those who work a night shift. Additionally, we observe a reverse pattern of how the negative arousal speech ratio changes in a work shift from Figure~\ref{fig:day_night_arousal}. These observations imply that night shift nurses are likely to express more positive and negative arousal at the start and middle of a work shift.

\section*{Speaking Pattern: ICU and Non-ICU units}

\subsection*{Speech Session}

\noindent \textbf{Statistical Analysis:} Here, we compare the speech activity pattern between ICU and non-ICU nurses. Exceptionally, we notice that shift pattern is a strong indicator of the session properties and can create substantial bias in comparing ICU and non-ICU nurses. Therefore, we compare the speaking patterns between these two work contexts to different shift patterns. From the results shown in Table~\ref{tab:session_icu_shift}, there are no differences in inter-session time and $>$1min Session \% between ICU nurses and non-ICU nurses. However, we can particularly identify that ICU nurses have a higher ratio of speech activity than non-ICU nurses in the patient room area. On the other hand, non-ICU nurses conduct most speech activities in the nursing station compared to ICU nurses. Moreover, these patterns are consistent in both day-shift and night-shift working schedules.

\begin{figure*}
    \centering
    \begin{tikzpicture}

        \node[draw=none,fill=none] at (0, 4){\includegraphics[width=0.5\linewidth]{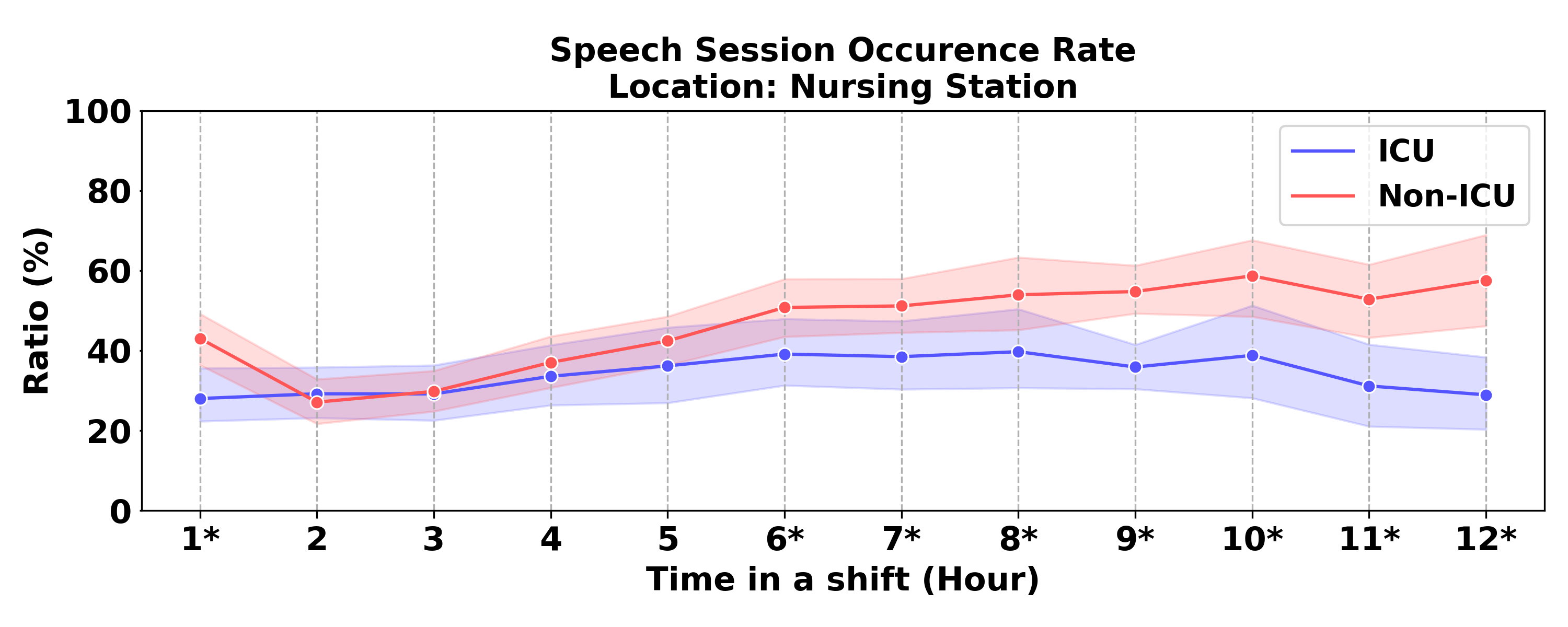}};

        \node[draw=none,fill=none] at (0.5\linewidth, 4){\includegraphics[width=0.5\linewidth]{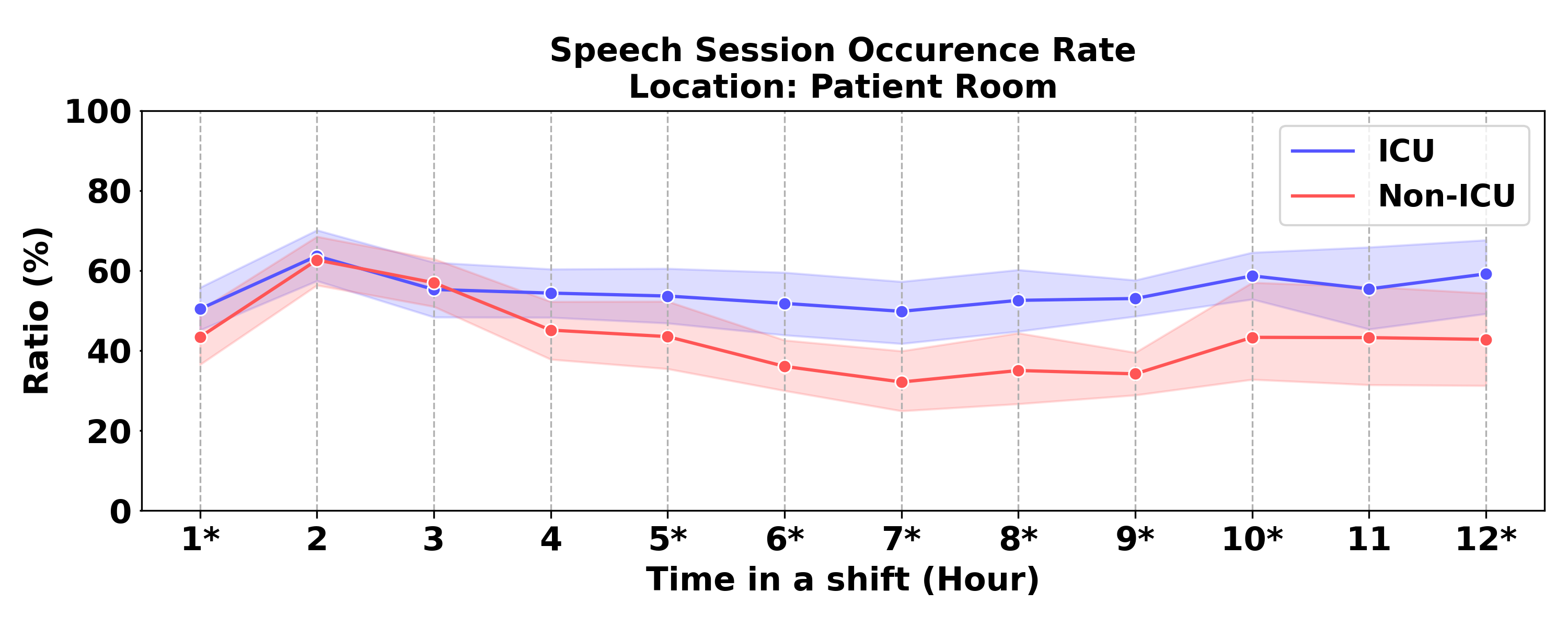}};
        
        \node[text width=3cm] at (0.3\linewidth, 5.75) {\textbf{Night shifts}};

        \node[draw=none,fill=none] at (0, 7.5){\includegraphics[width=0.5\linewidth]{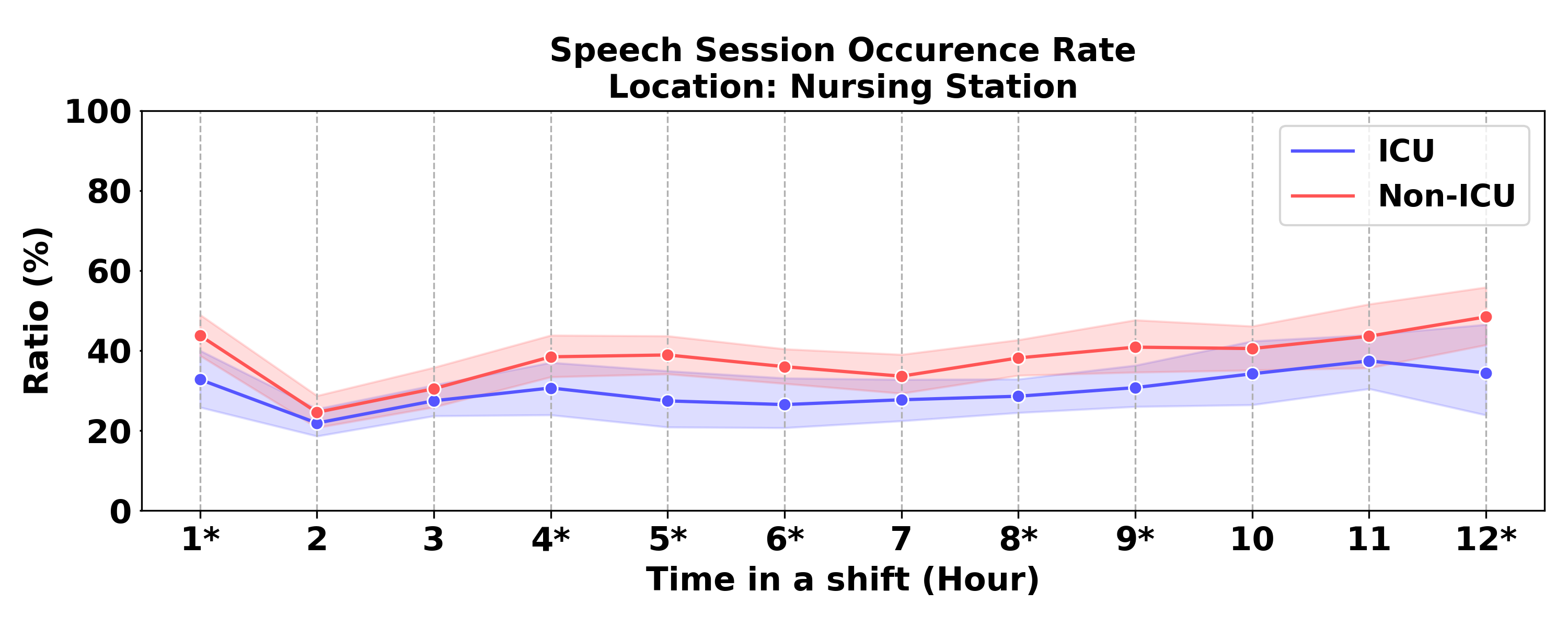}};

        \node[draw=none,fill=none] at (0.5\linewidth, 7.5){\includegraphics[width=0.5\linewidth]{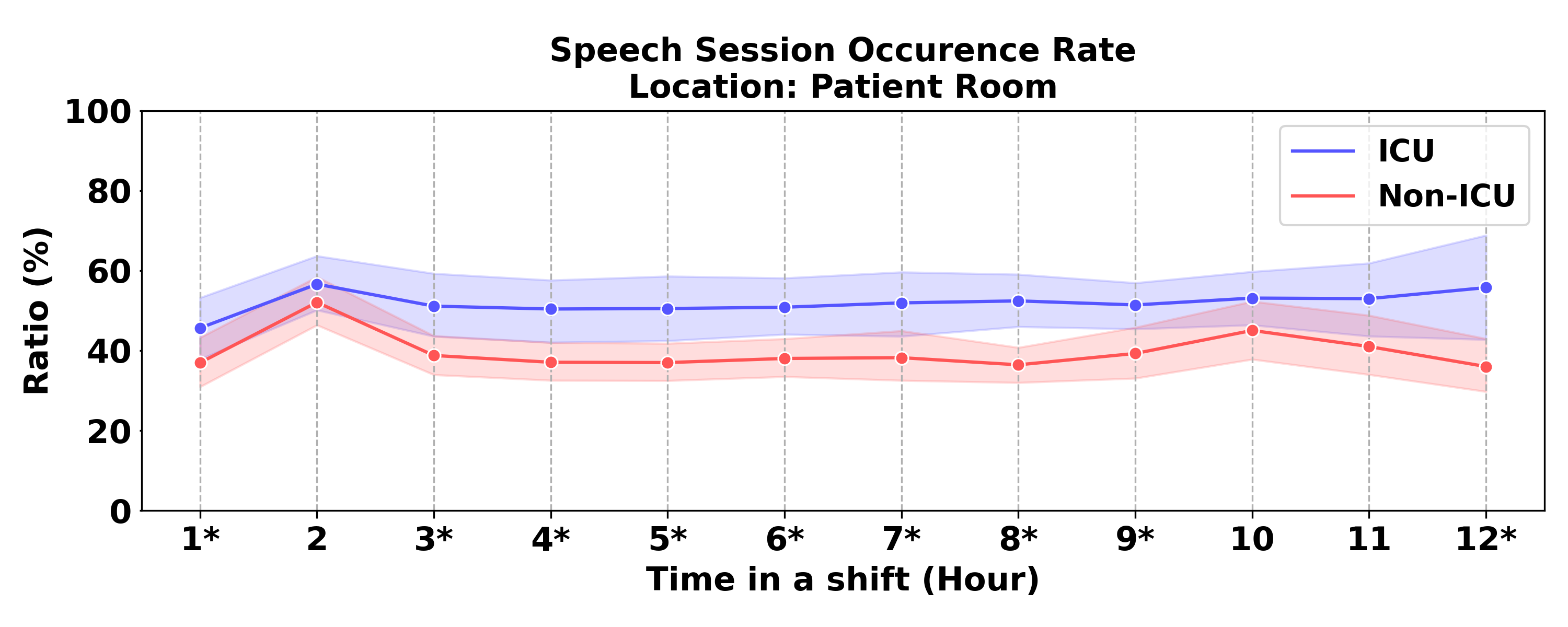}};

        \node[text width=3cm] at (0.3\linewidth, 9) {\textbf{Day shifts}};

    \end{tikzpicture}
    
    \vspace{-3.5mm}
    
    \caption{Comparison of the session occurrence rate (location) between ICU and non-ICU nurses. Statistical significance is performed using the Mann–Whitney U test: $\mathbf{p^{*}<0.05}$.}
    \label{fig:icu_sesssion}
    
\end{figure*}

\noindent \textbf{Dynamic Patterns:} The dynamic changes in session occurrence rate at different periods in a work shift are shown in Figure~\ref{fig:icu_sesssion}. Here, we focus on the analysis associated with the nursing station and the patient room area as these two speech session characteristics are substantially different, as shown in Table \ref{tab:session_icu_shift}. From Figure \ref{fig:icu_sesssion}, we can recognize that ICU nurses have a higher speech session ratio than non-ICU nurses in the patient room area at many periods in a work shift. Interestingly, we find this difference is most significant in the middle (between $4^{th}$ and $9^{th}$ hour) of a work shift. Moreover, we find that non-ICU nurses working the night shift have a significantly higher speech session ratio in the nursing station at the end of the work shift. However, this difference does not exist between ICU and non-ICU nurses working day shift schedules.

\subsection*{Speech Arousal}

\noindent \textbf{Statistical Analysis:} We compare the arousal levels of the speech activity between the ICU colleagues and the non-ICU colleagues in table~\ref{tab:session_icu_shift}. The results show that ICU nurses show a higher level of positive arousal speech. We can further discover that this difference is associated with positive-arousal speech in the nursing station. Specifically, ICU nurses have more speech activities with positive-arousal ratings than non-ICU nurses. ICU and non-ICU nurses have a similar level of positive-arousal speech in the patient room area. On the other hand, we can identify that ICU nurses are also likely to produce more negative-arousal speech than non-ICU nurses in nursing stations.

\begin{table*}
    \centering
    \caption{Table showing the inter-session time (in minutes), $>$1min session ratio (\%), and session occurrence rate (location) between ICU nurses and Non-ICU nurses in two shift patterns. Statistical significance is performed using the Mann–Whitney U test: $\mathbf{p^{*}<0.05}$. The median and the mean value of each feature are presented with the notation: median (mean).}
    \small
    \begin{tabular}{p{4cm}p{2cm}p{2cm}p{1.5cm}}

        \toprule
        \rule{0pt}{1ex}
        & 
        \multicolumn{3}{c}{Day shift}
        \rule{0pt}{1ex} \\
         \cmidrule(lr){2-4}
        
        \multirow{2}{*}{Feature} & 
        \multicolumn{1}{c}{ICU} & 
        \multicolumn{1}{c}{Non-ICU} &  
        \multicolumn{1}{c}{\multirow{2}{*}{{\centering p-value}}} \rule{0pt}{1ex} \\ 
        
        & 
        \multicolumn{1}{c}{(n=21)} &
        \multicolumn{1}{c}{(n=35)} & 
        \rule{0pt}{1ex} \\
        \cmidrule(lr){1-1} \cmidrule(lr){2-3} \cmidrule(lr){4-4}
        
        \multicolumn{1}{l}{\textbf{Inter-session Time (min)}} &
        \multicolumn{1}{c}{$5.06$ ($5.94$)} &
        \multicolumn{1}{c}{$5.89$ ($7.17$)} &
        \multicolumn{1}{c}{$0.099$}
        \rule{0pt}{2ex} \\

        \multicolumn{1}{l}{\textbf{$\mathbf{>}$1min Session \%}} & & &
        \rule{0pt}{1ex} \\

        \multicolumn{1}{l}{\hspace{0.5cm}{All Location}} &
        \multicolumn{1}{c}{$38.06$ ($37.83$)} &
        \multicolumn{1}{c}{$38.01$ ($35.81$)} &
        \multicolumn{1}{c}{$0.255$} 
        \rule{0pt}{2ex} \\
        
        \multicolumn{1}{l}{\hspace{0.5cm}{Nursing Station}} &
        \multicolumn{1}{c}{$30.52$ ($28.02$)} &
        \multicolumn{1}{c}{$30.53$ ($30.42$)} &
        \multicolumn{1}{c}{$0.220$}
        \rule{0pt}{2ex} \\
        
        \multicolumn{1}{l}{\hspace{0.5cm}{Patient Room}} &
        \multicolumn{1}{c}{$35.32$ ($33.85$)} &
        \multicolumn{1}{c}{$30.94$ ($31.31$)} &
        \multicolumn{1}{c}{$0.151$} 
        \rule{0pt}{2ex} \\

        \multicolumn{1}{l}{\textbf{Session Occurrence \%}} & & & \rule{0pt}{1ex} \\

        \multicolumn{1}{l}{\hspace{0.5cm}{Nursing Station}} &
        \multicolumn{1}{c}{$28.65$ ($26.76$)} &
        \multicolumn{1}{c}{$34.39$ ($33.80$)} &
        \multicolumn{1}{c}{$\mathbf{0.028^*}$} 
        \rule{0pt}{2ex} \\
        
        \multicolumn{1}{l}{\hspace{0.5cm}{Patient Room}} &
        \multicolumn{1}{c}{$52.66$ ($50.86$)} &
        \multicolumn{1}{c}{$39.79$ ($38.36$)} &
        \multicolumn{1}{c}{$\mathbf{<0.01^{*}}$}
        \rule{0pt}{2ex} \\
        
        \multicolumn{1}{l}{\hspace{0.5cm}{Lounge+Med.}} &
        \multicolumn{1}{c}{$8.29$ ($8.90$)} &
        \multicolumn{1}{c}{$13.50$ ($14.97$)} &
        \multicolumn{1}{c}{$\mathbf{<0.01^{*}}$} 
        \rule{0pt}{2ex} \\
        
        \multicolumn{1}{l}{\hspace{0.5cm}{Outside the Unit}} &
        \multicolumn{1}{c}{$13.23$ ($13.48$)} &
        \multicolumn{1}{c}{$12.21$ ($14.64$)} &
        \multicolumn{1}{c}{$0.349$} 
        \rule{0pt}{2ex} \\

        \multicolumn{1}{l}{\textbf{Positive Arousal \%}} & & & \rule{0pt}{1ex} \\
        
        \multicolumn{1}{l}{\hspace{0.5cm}{All Location}} &
        \multicolumn{1}{c}{$26.84$ ($30.29$)} &
        \multicolumn{1}{c}{$21.94$ ($22.14$)} &
        \multicolumn{1}{c}{$\mathbf{<0.01^{*}}$}
        \rule{0pt}{2ex} \\
        
        \multicolumn{1}{l}{\hspace{0.5cm}{Nursing Station}} &
        \multicolumn{1}{c}{$20.19$ ($27.00$)} &
        \multicolumn{1}{c}{$15.88$ ($16.34$)} &
        \multicolumn{1}{c}{$\mathbf{<0.01^{*}}$} 
        \rule{0pt}{2ex} \\
        
        \multicolumn{1}{l}{\hspace{0.5cm}{Patient Room}} &
        \multicolumn{1}{c}{$28.49$ ($32.15$)} &
        \multicolumn{1}{c}{$28.29$ ($26.32$)} &
        \multicolumn{1}{c}{$0.066$} 
        \rule{0pt}{2ex} \\

        \multicolumn{1}{l}{\textbf{Negative Arousal \%}} & & & \rule{0pt}{1ex} \\
        
        \multicolumn{1}{l}{\hspace{0.5cm}{All Location}} &
        \multicolumn{1}{c}{$24.20$ ($24.97$)} &
        \multicolumn{1}{c}{$27.76$ ($28.03$)} &
        \multicolumn{1}{c}{$\mathbf{0.012^*}$} 
        \rule{0pt}{2ex} \\
        
        \multicolumn{1}{l}{\hspace{0.5cm}{Nursing Station}} &
        \multicolumn{1}{c}{$23.92$ ($24.91$)} &
        \multicolumn{1}{c}{$31.78$ ($32.21$)} &
        \multicolumn{1}{c}{$\mathbf{<0.01^{*}}$}
        \rule{0pt}{2ex} \\
        
        \multicolumn{1}{l}{\hspace{0.5cm}{Patient Room}} &
        \multicolumn{1}{c}{$23.13$ ($23.78$)} &
        \multicolumn{1}{c}{$19.98$ ($21.56$)} &
        \multicolumn{1}{c}{$0.058$}
        \rule{0pt}{2ex} \\

        \midrule
        & 
        \multicolumn{3}{c}{Night shift}
        \rule{0pt}{1ex} \\
         \cmidrule(lr){2-4}
         
        \multirow{2}{*}{Feature} & 
        \multicolumn{1}{c}{ICU} & 
        \multicolumn{1}{c}{Non-ICU} &  
        \multicolumn{1}{c}{\multirow{2}{*}{{\centering p-value}}} \rule{0pt}{1ex} \\ 
        
        &
        \multicolumn{1}{c}{(n=21)} & 
        \multicolumn{1}{c}{(n=22)} & 
        \rule{0pt}{1ex} \\
        \cmidrule(lr){1-1} \cmidrule(lr){2-3} \cmidrule(lr){4-4}
        
        \multicolumn{1}{l}{\textbf{Inter-session Time (min)}} &
        \multicolumn{1}{c}{$8.51$ ($10.02$)} &
        \multicolumn{1}{c}{$9.00$ ($9.70$)} &
        \multicolumn{1}{c}{$0.456$}
        \rule{0pt}{2ex} \\

        \multicolumn{1}{l}{\textbf{$\mathbf{>}$1min Session \%}} & & &
        \rule{0pt}{1ex} \\

        \multicolumn{1}{l}{\hspace{0.5cm}{All Location}} &
        \multicolumn{1}{c}{$27.94$ ($30.09$)} &
        \multicolumn{1}{c}{$31.08$ ($31.42$)} &
        \multicolumn{1}{c}{$0.276$}
        \rule{0pt}{2ex} \\
        
        \multicolumn{1}{l}{\hspace{0.5cm}{Nursing Station}} &
        \multicolumn{1}{c}{$19.54$ ($19.65$)} &
        \multicolumn{1}{c}{$25.19$ ($26.59$)} &
        \multicolumn{1}{c}{$\mathbf{<0.01^{*}}$}
        \rule{0pt}{2ex} \\
        
        \multicolumn{1}{l}{\hspace{0.5cm}{Patient Room}} &
        \multicolumn{1}{c}{$25.72$ ($26.98$)} &
        \multicolumn{1}{c}{$28.32$ ($27.34$)} &
        \multicolumn{1}{c}{$0.418$}
        \rule{0pt}{2ex} \\

        \multicolumn{1}{l}{\textbf{Session Occurrence \%}} & & & \rule{0pt}{1ex} \\

        \multicolumn{1}{l}{\hspace{0.5cm}{Nursing Station}} &
        \multicolumn{1}{c}{$27.53$ ($30.61$)} &
        \multicolumn{1}{c}{$43.97$ ($41.81$)} &
        \multicolumn{1}{c}{$\mathbf{<0.01^{*}}$}
        \rule{0pt}{2ex} \\
        
        \multicolumn{1}{l}{\hspace{0.5cm}{Patient Room}} &
        \multicolumn{1}{c}{$56.22$ ($53.59$)} &
        \multicolumn{1}{c}{$38.74$ ($40.16$)} &
        \multicolumn{1}{c}{$\mathbf{<0.01^{*}}$}
        \rule{0pt}{2ex} \\
        
        \multicolumn{1}{l}{\hspace{0.5cm}{Lounge+Med.}} &
        \multicolumn{1}{c}{$7.49$ ($7.12$)} &
        \multicolumn{1}{c}{$8.18$ ($9.48$)} &
        \multicolumn{1}{c}{$0.269$}
        \rule{0pt}{2ex} \\
        
        \multicolumn{1}{l}{\hspace{0.5cm}{Outside the Unit}} &
        \multicolumn{1}{c}{$9.12$ ($11.07$)} &
        \multicolumn{1}{c}{$7.52$ ($8.69$)} &
        \multicolumn{1}{c}{$0.101$}
        \rule{0pt}{2ex} \\

        \multicolumn{1}{l}{\textbf{Positive Arousal \%}} & & & \rule{0pt}{1ex} \\
        
        \multicolumn{1}{l}{\hspace{0.5cm}{All Location}} &
        \multicolumn{1}{c}{$30.57$ ($29.40$)} &
        \multicolumn{1}{c}{$22.48$ ($20.36$)} &
        \multicolumn{1}{c}{$\mathbf{<0.01^{*}}$}
        \rule{0pt}{2ex} \\
        
        \multicolumn{1}{l}{\hspace{0.5cm}{Nursing Station}} &
        \multicolumn{1}{c}{$23.24$ ($24.17$)} &
        \multicolumn{1}{c}{$14.46$ ($14.82$)} &
        \multicolumn{1}{c}{$\mathbf{<0.01^{*}}$}
        \rule{0pt}{2ex} \\
        
        \multicolumn{1}{l}{\hspace{0.5cm}{Patient Room}} &
        \multicolumn{1}{c}{$32.09$ ($32.11$)} &
        \multicolumn{1}{c}{$26.40$ ($26.38$)} &
        \multicolumn{1}{c}{$\mathbf{0.048^*}$}
        \rule{0pt}{2ex} \\

        \multicolumn{1}{l}{\textbf{Negative Arousal \%}} & & & \rule{0pt}{1ex} \\
        
        \multicolumn{1}{l}{\hspace{0.5cm}{All Location}} &
        \multicolumn{1}{c}{$26.13$ ($26.31$)} &
        \multicolumn{1}{c}{$30.76$ ($35.84$)} &
        \multicolumn{1}{c}{$\mathbf{<0.01^{*}}$}
        \rule{0pt}{2ex} \\
        
        \multicolumn{1}{l}{\hspace{0.5cm}{Nursing Station}} &
        \multicolumn{1}{c}{$28.09$ ($28.43$)} &
        \multicolumn{1}{c}{$41.28$ ($42.68$)} &
        \multicolumn{1}{c}{$\mathbf{<0.01^{*}}$}
        \rule{0pt}{2ex} \\
        
        \multicolumn{1}{l}{\hspace{0.5cm}{Patient Room}} &
        \multicolumn{1}{c}{$23.61$ ($23.71$)} &
        \multicolumn{1}{c}{$24.01$ ($27.44$)} &
        \multicolumn{1}{c}{$0.476$}
        \rule{0pt}{2ex} \\
        
        \bottomrule

    \end{tabular}
    
    \vspace{-3.5mm}
    \label{tab:session_icu_shift}
\end{table*}

\begin{figure*}
    \centering
    \begin{tikzpicture}

        \node[draw=none,fill=none] at (0, 4){\includegraphics[width=0.5\linewidth]{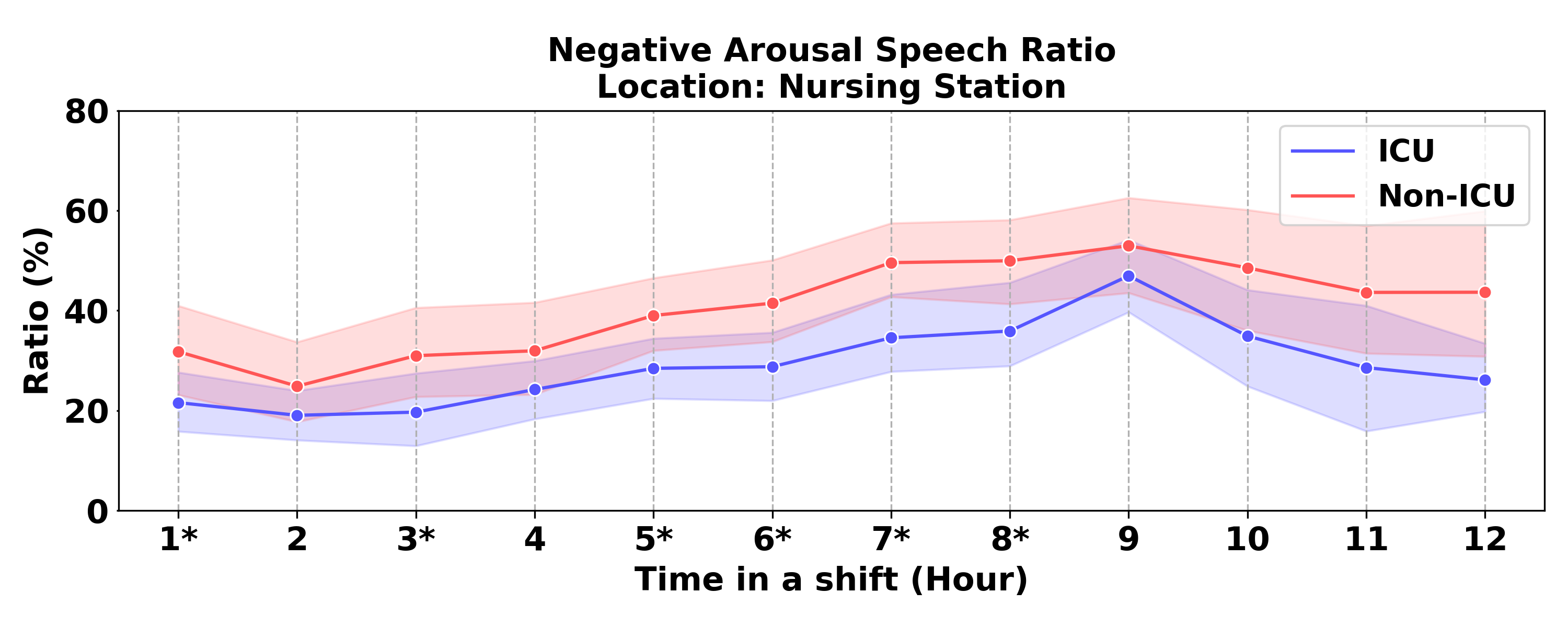}};

        \node[draw=none,fill=none] at (0.5\linewidth, 4){\includegraphics[width=0.5\linewidth]{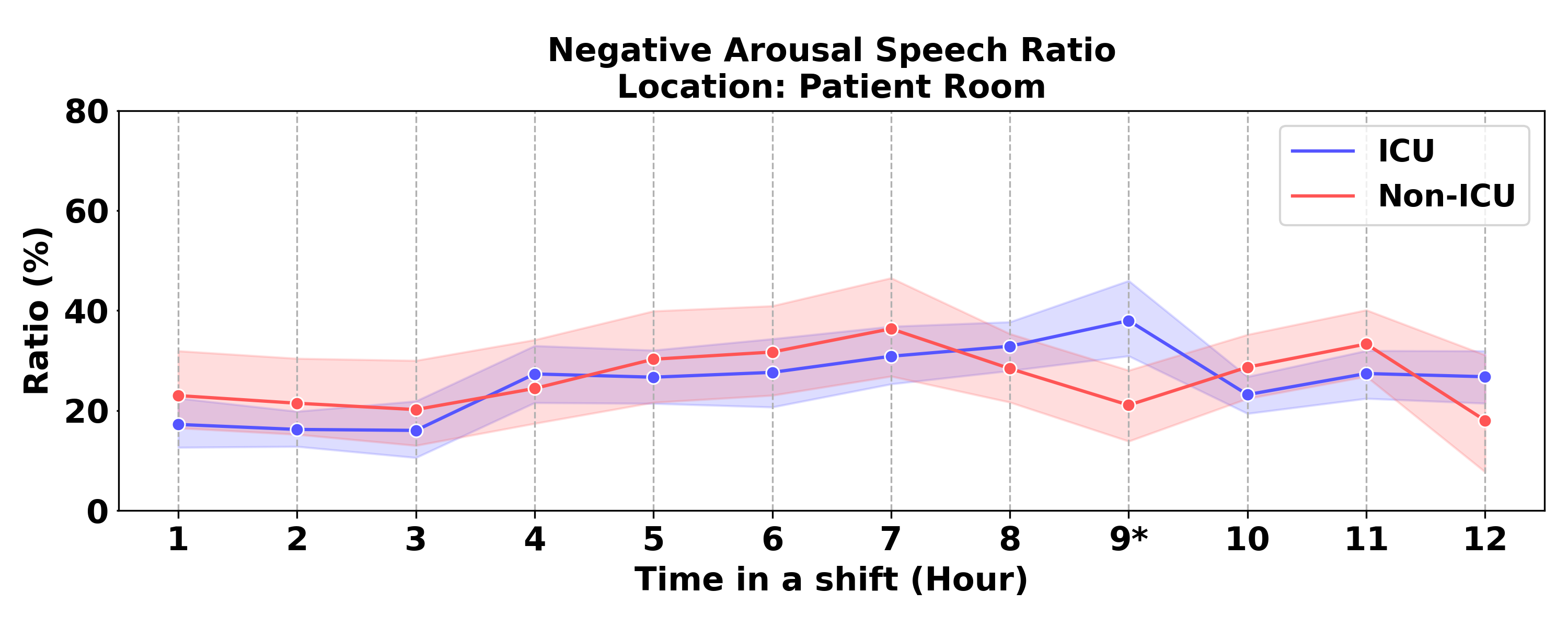}};
        
        \node[text width=3cm] at (0.3\linewidth, 5.75) {\textbf{Night shifts}};

        \node[draw=none,fill=none] at (0, 7.5){\includegraphics[width=0.5\linewidth]{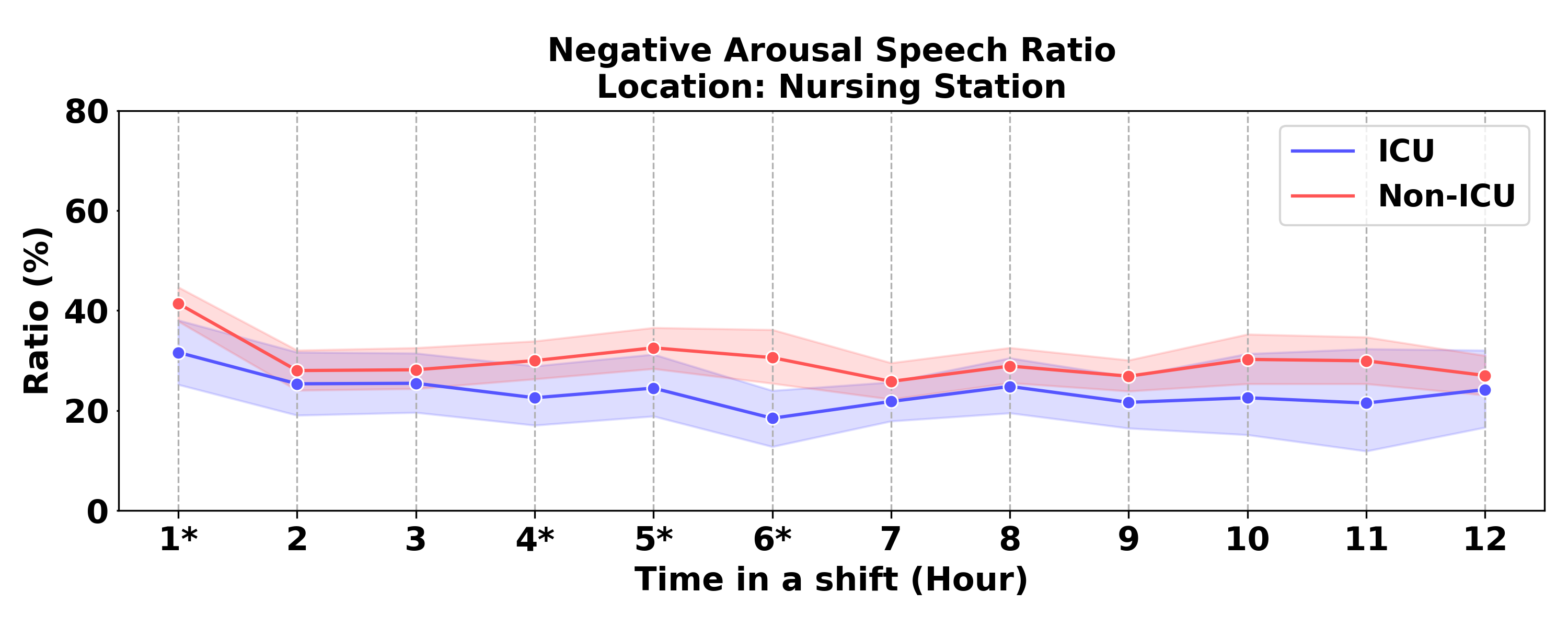}};

        \node[draw=none,fill=none] at (0.5\linewidth, 7.5){\includegraphics[width=0.5\linewidth]{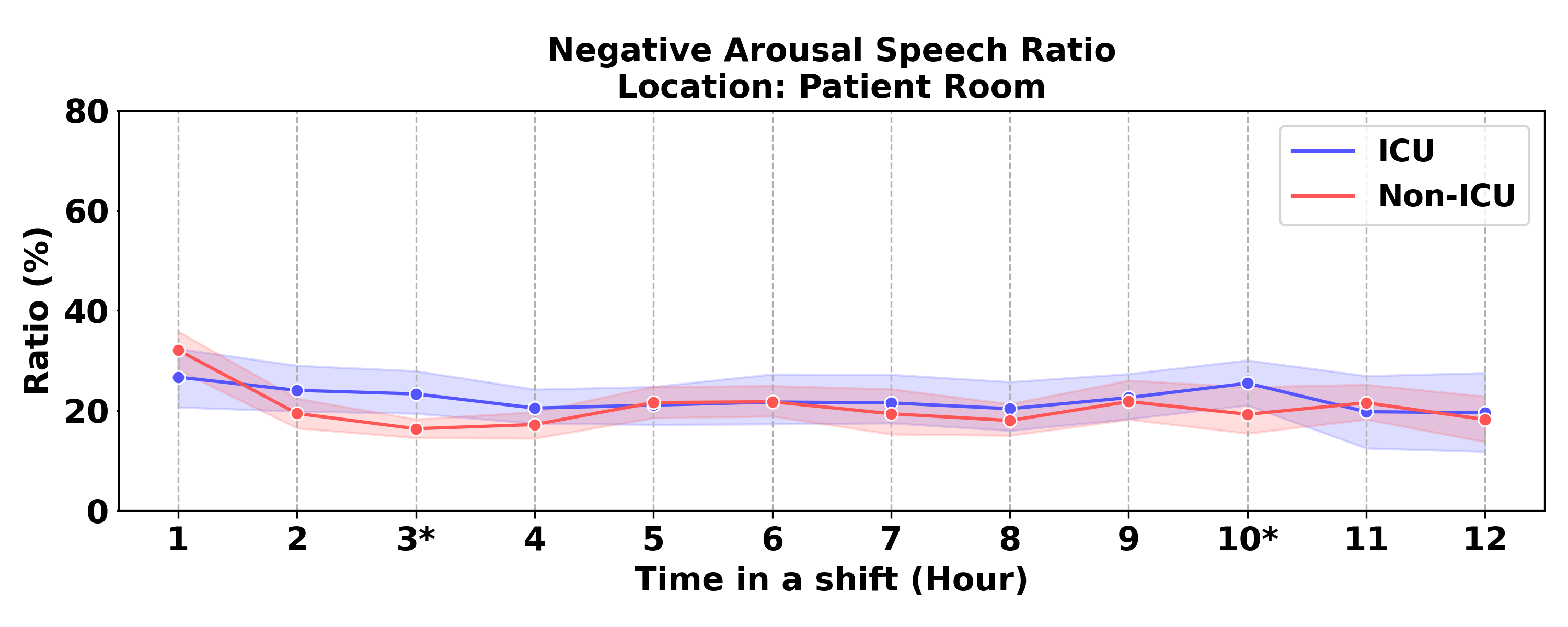}};

        \node[text width=3cm] at (0.3\linewidth, 9.0) {\textbf{Day shifts}};

    \end{tikzpicture}
    
    \vspace{-2.5mm}
    
    \caption{Comparison of the negative-arousal speech activity ratio between ICU and non-ICU nurses. Statistical significance is performed using the Mann–Whitney U test: $\mathbf{p^{*}<0.05}$.}
    \label{fig:icu_arousal}
    
\end{figure*}

\noindent \textbf{Dynamic Patterns:} In this analysis, we show the comparisons of negative-arousal speech between nurses in an ICU unit and those in a non-ICU unit in Figure~\ref{fig:icu_arousal}. Similar to the analysis conducted in the last section, we perform a comparison by controlling the shift pattern variable. Firstly, we observe that ICU nurses who work a day shift exhibit a higher negative-arousal speech ratio than non-ICU nurses in the patient room area for the majority of the working shift. Interestingly, we also observe that the negative-arousal speech ratio associated with the patient rooms decreases substantially at the end of the working shift among the ICU nurses who work the day shift schedule. Although ICU nurses who work the night shift schedule show higher levels of negative arousal speech ratio in the nursing station than non-ICU nurses, we find that this difference is only significant at the beginning and also around the end of the work shift.

\section*{Speaking Pattern: Machine Learning Validation}

In this section, we conduct a Machine Learning experiment that uses the speech activity features to predict the self-reported baselines described in section~\ref{sec:study_design}-\ref{sec:baseline_variables}.

\subsection*{Experiment Details}
In particular, we apply 2 statistical functionals (mean and average) to the extracted speaking pattern features, including inter-session time, above 1min session ratio, and positive/negative arousal speech ratio. In addition, as we are interested in studying the location-context of the speech activity, we also apply the statistical functionals to the speech session occurrence ratio, the positive/negative arousal speech ratio, and the above 1min session ratio in the nursing station and the patient rooms. We choose these two locations since nurses spend most of their time in work shifts in these two locations. Meanwhile, the speaking patterns, such as speech arousal, change from the start to the end of a working shift. Therefore, we decided to extract the positive/negative arousal speech ratio at the start, middle, and end of the work shift. 

In addition to the speaking patterns, we extracted a set of physiological features using the data recorded from the Fitbit wristband. In particular, we follow our previous work \cite{feng2021multimodal} to extract the daily walk activity ratio and sleep duration features. We compute the average and standard deviation for each measure. We also add the shift and the working unit as the input feature for the machine learning experiment since behavior patterns are substantially different between distinct shift patterns and working units. On the other hand, we binarize the self-report variables using the median value of each measure to create a balanced label distribution. Finally, we investigated the efficacy of the Random Forest (RF) classifier trained using the physiological and speech activity features we extracted above. We perform z-normalization on each feature to remove scaling differences. We apply a 5-fold cross-validation scheme combined with grid search to evaluate the performance of our proposed classifier. We report the best results measured in micro-F1 scores.

\begin{table}
    \centering
    \caption{Table showing the prediction of self-reported baselines using Random Forest with speech activity features and physiological features.}

    \begin{tabular}{p{2cm}p{1.5cm}p{1.5cm}p{1.5cm}}

        \toprule
        \multicolumn{1}{c}{Baseline} & 
        \multicolumn{1}{c}{Physiological} & 
        \multicolumn{1}{c}{Speaking Pattern} &  
        \multicolumn{1}{c}{Multi-modal} \\
        \multicolumn{1}{c}{Measures} & 
        \multicolumn{1}{c}{Features} & 
        \multicolumn{1}{c}{Features} & 
        \multicolumn{1}{c}{Features} \\
        \cmidrule(lr){1-1} \cmidrule(lr){2-2} \cmidrule(lr){3-3} \cmidrule(lr){4-4}
        
        \multicolumn{1}{l}{{Postive Affect}} &
        \multicolumn{1}{c}{$50.52\%$} &
        \multicolumn{1}{c}{${51.47\%}$} &
        \multicolumn{1}{c}{$\mathbf{51.57\%}$}  \\

        \multicolumn{1}{l}{{Negative Affect}} &
        \multicolumn{1}{c}{$64.68\%$} &
        \multicolumn{1}{c}{$65.58\%$} &
        \multicolumn{1}{c}{$\mathbf{65.68\%}$}  \\

        \multicolumn{1}{l}{{Life Satisfaction}} &
        \multicolumn{1}{c}{$54.47\%$} &
        \multicolumn{1}{c}{${56.63\%}$} &
        \multicolumn{1}{c}{$\mathbf{57.63\%}$}  \\

        \bottomrule

    \end{tabular}
    
    \label{tab:ml_result}
    \vspace{-3mm}
\end{table}

\subsection*{Results}
\begin{figure}
    \centering
    \begin{tikzpicture}
        
        \node[draw=none,fill=none] at (0, 4){\includegraphics[width=0.75\linewidth]{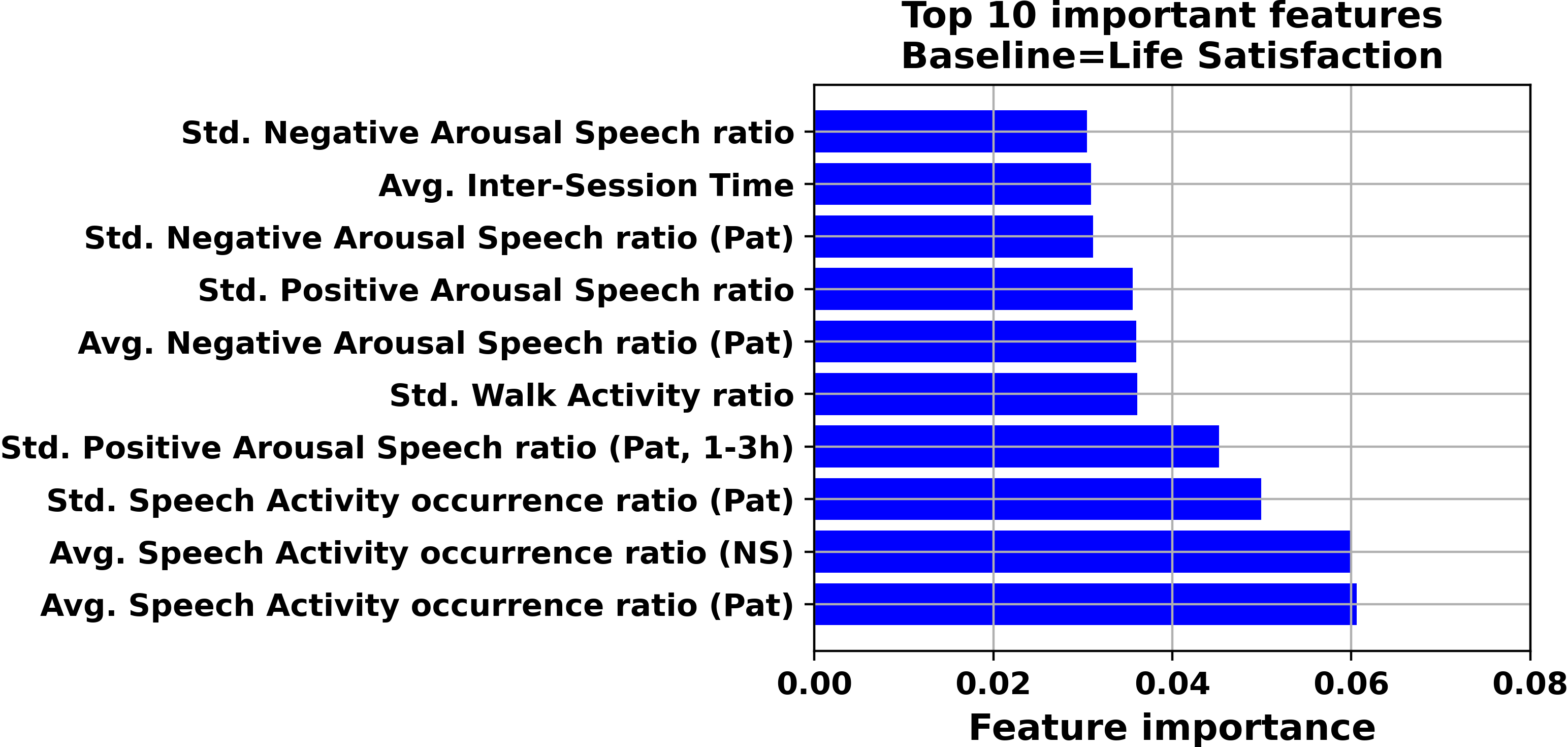}};
        
        \node[draw=none,fill=none] at (0, 8){\includegraphics[width=0.75\linewidth]{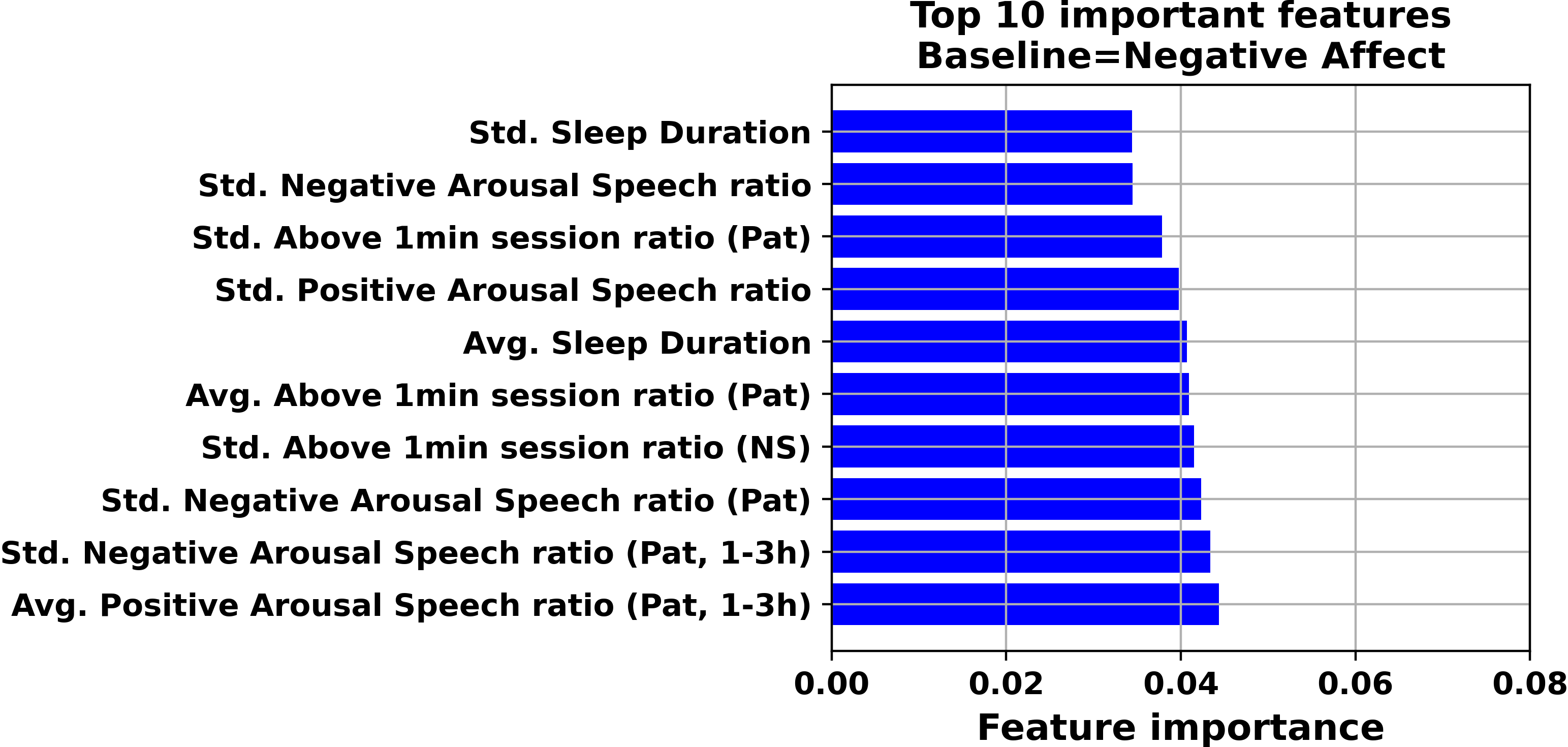}};

        \node[draw=none,fill=none] at (0, 12){\includegraphics[width=0.75\linewidth]{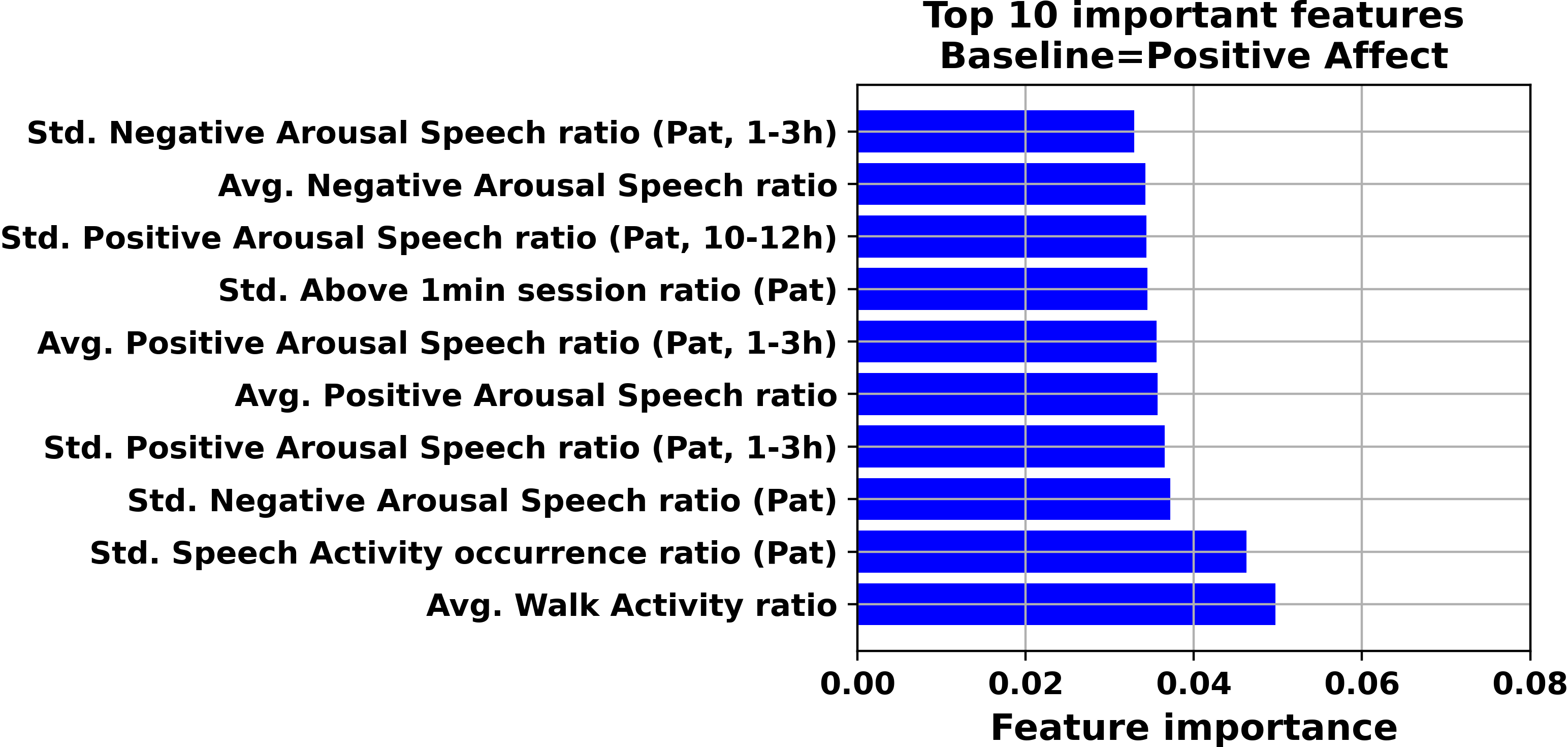}};

    \end{tikzpicture}
    
    \caption{Top 10 important features for predicting self-reported positive affect, negative affect, and life satisfaction scores using Random Forest Model}
    \label{fig:feature_importance}
    
\end{figure}

Table~\ref{tab:ml_result} compares the best prediction results (on the F1 score) of the Random Forest classifier among speaking pattern features, physiological features, and multi-modal features. The results indicate that our best prediction results on life satisfaction scores are achieved using features associated with speaking patterns. Moreover, the multi-modal features yield the best result to predict the positive and negative affect scores. To better understand our model, we report the top 10 most important features in our Random Forest model in Figure~\ref{fig:feature_importance}. We notice that speaking patterns are consistently among the top 3 most important features in the Random Forest model. Increasingly, features associated with negative-arousal speech activity ratio are rated as the first, third, and 4th important features in predicting positive affect, negative affect, and life satisfaction. These findings demonstrate that the extracted speaking patterns can be used to improve the predictions for affect measures and life satisfaction.

\section*{Conclusion}

In this work, we study the speaking patterns of nursing professionals using recorded indoor proximity data and foreground speech characteristics measures. The wearable sensors and devices we used in this data analysis include TARs (TILES audio recorder) and Bluetooth hubs installed throughout the hospital environment. Our data were collected from nursing professionals working in a complex hospital environment across a 10 week period. Our multi-modal processing pipeline includes indoor location estimation and speaking pattern extraction. We first infer the foreground speech activity using a pre-trained deep-learning model and further extract the speaking patterns from foreground speech characteristics. Moreover, we compute location-based speaking patterns by combing indoor location estimation and foreground speech characteristics for post-analysis.

Our analysis reveals that day-shift nurses engage in more frequent and prolonged speech activities than night-shift nurses. Nurses who work the day shift also have more speech activity in the location of the nursing station. Moreover, night shift nurses show higher levels of negative arousal in speech activity than day shift nurses. We further compare the speaking patterns between nurses working in the ICU and non-ICU units. We find that non-ICU nurses show more speech activity in the nursing stations. In comparison, ICU nurses have more speech activity in the patient room area. Similarly, we can identify that ICU nurses have more positive arousal speech activity in the nursing station. The dynamic patterns of speaking patterns show that night shift nurses have fewer speech sessions, shorter speech sessions, and more negative arousal speech activity in the middle of the work shift compared to the start and the end of the work shift. On the other hand, non-ICU nurses have fewer speech activity in the patient room areas, particularly in the middle of the work shift. Lastly, our machine learning experiments show that the speaking pattern features with daily physiological features can also predict affect and life satisfaction measures.

These findings show promises of using wearable technologies to study speaking patterns unobtrusively in complex real-world contexts such as workplaces. The results also demonstrate that our proposed multi-modal speaking modeling pipeline can extract unique speaking patterns among people with different work schedules and environments.

\section*{Acknowledgement}
The research is based upon work supported by the Office of the Director of National Intelligence(ODNI), Intelligence Advanced Research Projects Activity(IARPA), via IARPA Contract No 2017 - 17042800005. The views and conclusions contained herein are those of the authors and should not be interpreted as necessarily representing official policies or endorsements, either expressed or implied, of ODNI, IARPA, or U.S. Government. The U.S. Government is authorized to reproduce and distribute reprints for Governmental purposes notwithstanding any copyright annotation thereon.

\nolinenumbers

\bibliography{ref}

\end{document}